\documentclass[traditabstract]{aa}
\usepackage{graphicx}
\usepackage{txfonts}
\usepackage{amsmath}
\usepackage{amsfonts}
\usepackage{amssymb}
\usepackage{booktabs}
\usepackage{setspace}
\usepackage[colorlinks,citecolor=blue]{hyperref}
\usepackage{blindtext}
\usepackage[english]{babel}
\usepackage{natbib}
\usepackage{tabularx,multirow,booktabs,blindtext}
\usepackage{subfigure}
\usepackage{lipsum}
\usepackage{adjustbox}
\usepackage{longtable}
\usepackage{rotating} 
\usepackage{lscape}
\usepackage{footnote}
\usepackage{threeparttable}
\usepackage{multicol}
\usepackage{graphicx}
\usepackage{caption}
\usepackage{tablefootnote}

\newcommand*{\logten}{\mathop{\log_{10}}}
\bibpunct{(}{)}{;}{a}{}{,}

\begin{document}
        
\title{Relation of X-ray activity and rotation in M dwarfs\\ 
        and predicted time-evolution of the X-ray luminosity\thanks{The collection of all 
                updated data from the literature is listed in Table~\ref{ALLdata_par}, 
                also available in electronic form
                at the CDS via anonymous ftp to cdsarc.u-strasbg.fr (130.79.128.5)
                or via http://cdsweb.u-strasbg.fr/cgi-bin/qcat?J/A+A/.}}
\titlerunning{Relation of X-ray activity to rotation in M dwarfs}
\authorrunning{Magaudda et al.}
\author{E. Magaudda$^1$, B. Stelzer$^{1,2}$, K. R. Covey$^3$, St. Raetz$^1$, S. P. Matt$^4$, and A.Scholz$^5$}
\institute{ Institut f\"{u}r Astronomie und Astrophysik, Eberhard-Karls Universit\"{a}t T\"{u}bingen, Sand 1, D-72076 T\"{u}bingen, Germany
\and INAF- Osservatorio Astronomico di Palermo, Piazza Parlamento 1, 90134 Palermo, Italy
\and Department of Physics \& Astronomy, Western Washington University, Bellingham WA 98225-9164, USA
\and University of Exeter, Department of Physics \& Astronomy, Physics Bldg., Stocker Road, Exeter EX4 4QL, UK
\and SUPA, School of Physics \& Astronomy, University of St Andrews, North Haugh, St Andrews, KY169SS, UK}
\date{Received 24 December 2019 / Accepted 30 March 2020}
\abstract
        {The relation of activity to rotation in M dwarfs is of high astrophysical interest because it provides observational evidence of the stellar dynamo, which is poorly understood for low-mass stars, especially in the fully convective regime.
        Previous studies have shown that the relation of X-ray activity to rotation consists of two different regimes: the \textit{\textup{saturated}} regime for fast-rotating stars and the \textit{\textup{unsaturated}} regime for slowly rotating stars. The transition between the two regimes lies at a rotation period of $\sim10$\,d. We present here a sample of 14 M dwarf stars observed with {\em XMM-Newton} and {\em Chandra}, for which we also computed rotational periods from {\em Kepler Two-Wheel} ({\em K2}) Mission light curves. We compiled X-ray and rotation data from the literature and homogenized all data sets to provide the largest uniform sample of M dwarfs (302 stars) for X-ray activity and rotation studies to date. We then fit the relation between $L_{\rm x} - P_{\rm rot}$ using three different mass bins to separate partially and fully convective stars. We found a steeper slope in the unsaturated regime for fully convective stars and a nonconstant $L_{\rm x}$ level in the saturated regime for all masses. In the $L_{\rm x}/L_{\rm bol}-R_{\rm O}$ space we discovered a remarkable double gap that might be related to a discontinuous period evolution.
        Then we combined the evolution of $P_{\rm rot}$ predicted by angular momentum evolution models with our new results on the empirical $L_{\rm x} - P_{\rm rot}$ relation to provide an estimate for the age decay of X-ray luminosity. We compare predictions of this relationship with the actual X-ray luminosities of M stars with known ages from 100~Myr to a few billion years. We find remarkably good agreement between the predicted $L_{\rm x}$ and the observed values for partially convective stars. However, for fully convective stars at ages of a few billion years, the constructed $L_{\rm x}-$age relation overpredicts the X-ray luminosity because the angular momentum evolution model underpredicts the rotation period of these stars. Finally, we examine the effect of different parameterizations for the Rossby number ($R_{\rm O}$) on the shape of the activity-rotation relation in $L_{\rm x}/L_{\rm bol}-R_{\rm O}$ space, and we find that the slope in the unsaturated regime and the location of the break point of the dual power-law depend sensitively on the choice of $R_{\rm O}$. }
   

\maketitle
\section{Introduction}
\label{intro}
        Late-type stars emit X-rays from their outermost atmospheric layer. This is called the corona\textup{\textup{}}. The layer consists of a magnetically confined plasma with temperatures of up to several million Kelvin. The stellar corona was first observed in the Sun and is thought to be heated by the release of magnetic energy through a dynamo mechanism. For G-type stars, convection in the outer envelope together with differential rotation generates magnetic activity through an $\alpha\Omega$-dynamo mechanism \citep{Parker:1955aa}.
        The amount of magnetic energy that is released in the corona decreases over the stellar lifetime. This is a result of rotational spin-down that leads to decreased dynamo efficiency. The spin-down is driven by mass loss that interacts with the magnetic field. Stellar rotation and magnetism thus form a complex feedback system. 

        From the observations of the Sun, it is known that activity has distinct observable manifestations in each atmospheric layer. The photosphere\textup{} on a magnetically active star contains regions that are cooler and darker than their surroundings, called star spots\textit{}. In these regions the magnetic pressure is so high that it overcomes the gas pressure and consequently inhibits the heat transport by convection. As an observable consequence, the light curve displays a periodic brightness modulation caused by the rotation of the dark spots \citep{Eaton:1979aa, Bopp1973}. With space-based missions such as {\em Kepler} and its successor {\em K2}, we can therefore measure the stellar rotation period ($P_{\rm rot}$) and detect the photometric variations because the magnetically active regions continuously cross the visible hemisphere as the star rotates.
        The outer two atmospheric layers, the chromosphere\textit{} and the corona\textit{}, can be analyzed with optical, UV, and X-ray observations \citep{Guedel2004,Durney1993}. 
                 
        Main-sequence stars with $M_{\star}\lesssim0.35\,\rm {M_{\odot}}$ \citep{Chabrier2006}, corresponding roughly to spectral types equal to or later than M3.5, have fully convective interiors. They therefore lack the tachocline  observed in solar-type stars, which invalidates the $\alpha\Omega$-dynamo mechanism. Possible alternative magnetic processes are an $\alpha^2$-dynamo or a turbulent dynamo mechanism, for which the dependence on rotation has not yet been settled.
        
        An observational way to indirectly examine the underlying stellar dynamo that causes the magnetic activity in solar- and later-type stars is to study the relation of coronal activity to rotation. This relation is typically expressed in terms of the X-ray luminosity ($L_{\rm x}$) as a function of the rotational period ($P_{\rm rot}$), or alternatively, in terms of the ratio between the stellar X-ray and bolometric luminosities ($\frac{L_{\rm x}}{L_{\rm bol}}$) as a function of the Rossby number{\em } ($R_{\rm O}$). This variable is a dimensionless number defined as the ratio between $P_{\rm rot}$ and the convective turnover time ($\tau_{\rm conv}$), the time needed for a convective cell to rise to the surface. Because $\tau_{\rm conv}$ is not an observable parameter, the use of $R_{\rm O}$ introduces a model dependence or requires an ad hoc description.
        
        Previous studies of the activity-rotation relation have shown two different regimes. In particular, for fast-rotating stars, the X-ray activity does not depend on the rotation (saturated regime), while on the other hand, the X-ray activity of slowly rotating stars declines with increasing rotational period (unsaturated regime). 
        \citet{Pallavicini1981} were the first to study the coronal X-ray emission as a function of  $v \sin i$ for a sample of stellar spectral type (O3 to M). Later, \citet{Pizzolato2003} studied the coronal X-ray emission and stellar rotation in late-type main-sequence stars with X-ray data from the {\em ROSAT} satellite and calculated $P_{\rm rot}$ from $v \sin i$ measurements. All $v \sin i$ values have been translated into $P_{\rm rot}$ upper limits because of the unknown inclination of the stellar systems, and only two M dwarf stars were located in the unsaturated regime. Therefore the relation remained poorly constrained.
        Since then, \citet{Wright2011,Wright2016,Stelzer2016,Wright2018} and \citet{Gonzalez-Alvarez2019} have studied the X-ray activity-rotation relation of M dwarfs based on photometric $P_{\rm rot}$, collecting much more information about the empirical connection between rotation and X-ray emission.

        Because stellar rotation slows down throughout the main-sequence life of a star, the dynamo efficiency also decreases over time. This entails a decrease in X-ray luminosity. The joint evolution of rotation and activity is encoded in the empirical rotation-activity relation. Direct observations of the age decay of rotation and X-ray emission are hampered for M dwarfs by the lack of stars with known age. Direct observations of the age decay of rotation and X-ray emission are difficult to obtain for M dwarfs with known ages of 1 Gyr or older. The availability of precise light curves from the {\em Kepler}/K2 mission, coupled with targeted or serendipitous X-ray observations, has enabled detailed studies of the relation of age, rotation, and activity in several $\sim$600\,Myr benchmark open clusters (i.e., Praesepe \& Hyades, \citealt{Douglas2014},\citealt{Nunez2015}; M37, \citealt{Nunez2017}),  but the rotation periods and activity measures required to calibrate models of angular momentum evolution are only now becoming available for M dwarfs in clusters older than 1 Gyr \citep[i.e., NGC~752;][]{Agueros2018}. For this reason, angular momentum evolution models for M dwarfs \citep{Matt_2015} have not been calibrated for stars beyond the ages of the Hyades ($\sim 600$\,Myr). Spin-down models can be used combined with the empirical rotation-activity relation to predict the long-term evolution of stellar X-ray emission, however.
        
        In this work, we present an updated relation of  X-ray activity to rotation in M dwarf stars and predict their $L_{\rm x}-$age relation.
        In Sect.~\ref{sample} we introduce the sample of M dwarfs that we studied, which includes new X-ray observations from {\em Chandra} and {\em XMM-Newton,} and new rotation periods from the {\em K2} mission as well as a collection of the samples studied in the previous literature. In Sect.~\ref{stellar_par} we describe how we derived the stellar parameters and how we updated the literature sample to provide the largest and most homogeneous database to date for studies of M-dwarf rotation and coronal activity. In Sect.~\ref{data_analysis} we describe our analysis of the new {\em XMM-Newton} and {\em Chandra} observations, and in Sect.~\ref{K2_an} we present our selection and study of the rotation periods derived for the stars with new X-ray data. Sect.~\ref{results} contains the results of the observed relation of rotation to activity in terms of 
        $L_{\rm x} - P_{\rm rot}$ and $L_{\rm x}/L_{\rm bol} - R_{\rm O}$, and our construction of the $L_{\rm x} - $age relation with help of the spin-down models. A summary and discussion of our results is presented in Sect.~\ref{disc}, followed by our conclusions and the outlook for further development in Sect.~\ref{Concl}. 
\section{Sample selection}
        \label{sample}
        We observed 14 M-dwarf stars with {\em XMM-Newton} or {\em Chandra}. The sample was extracted from the stars of the Superblink proper motion catalog by \citet{L_pine_2011} (henceforth LG11) that have {\em K2} rotation period measurements. The LG11 catalog is an all-sky list of 8889 M dwarfs (SpT= K7 to M6) brighter than $J = 10$~mag and within 100~pc. Rotation periods have been determined for the LG11 stars in the {\em K2} fields of campaigns C0 to C4 by \citet{Stelzer2016}, and the periods of the LG11 stars located in successive {\em K2} campaigns were measured by us using the same methods (Raetz et al. 2020, AN~subm.). For the X-ray observations obtained for the present study, we predominantly selected stars with long rotation periods ($P_{\rm rot} > 10$~d).
        Our new {\em XMM-Newton} and {\em Chandra} sample covers periods of 0.6 to 79~d.
        
        Together with the 14 new stars, we here present the whole sample from the previous literature on the X-ray activity-rotation relation of M dwarfs based on photometric periods, that is, \citet{Wright2011,Wright2016,Stelzer2016,Wright2018}, and \citet{Gonzalez-Alvarez2019}. The total sample we consider consists of 302 M~dwarfs.
        To obtain a homogeneous sample, we applied some updates to the parameters of the stars from the literature. In the next section we describe our updating procedure together with the determination of the stellar parameters for our new sample of 14 stars with {\em K2} rotation periods and deep X-ray observations.
\section{Sample properties}
\label{stellar_par}
        In this section we explain the method we used to compute distances and stellar parameters for the 14 new X-ray observations (henceforth ``our sample") and for all literature samples we list in Sect.~\ref{sample}. Throughout the paper we refer to the ``full sample" when we consider the 14 new observations together with the 288 stars from the literature.
        First, we evaluated {\em Gaia-DR2} parallaxes to obtain updated distances (henceforth $\rm d_{gaia}$).
        {\em Gaia-DR2} contains spurious astrometric solutions \citep{Arenou2018}, therefore it is important to consider quality flags. To do so, we examined the available {\em Gaia} parallaxes for all samples using the filters provided by \citet{Lindegren2018} in their Appendix C. 
        For the stars without {\em Gaia} parallax or stars that are not validated by the quality flags, we calculated photometric distances (henceforth $\rm d_{phot}$). To this end, we made use of photometric magnitudes from the USNO CCD Astrograph (UCAC4)\footnote{We verified that the UCAC4 $\rm V_{mag}$ are reliable by comparing them with the {\em Gaia}-to-$\rm V_{mag}$ conversion from \citet{Jao_2018}.} and the Two Micron All-Sky Survey (2MASS) catalogs 
         and applied the empirical relation from \citet{Stelzer2016} to calculate the absolute magnitude in the $K$ band ($M_{\rm K_{s}}$) from $V-J$. We then used $M_{\rm K_{s}}$ together with the observed apparent magnitude in $K$ band ($\rm K_{s}$) to derive the photometric distances. When we compared the two distance estimates, we identified 37 stars ($\sim 12\%$) for which $d_{\rm Gaia}\geq 2\cdot d_{\rm phot}$, and for these cases, we adopt the photometric distance throughout. There are no stars for which the Gaia distance is significantly smaller than the photometric distance.
        The $FLAG_ {\rm D}$ column in Tables~\ref{Par} and \ref{ALLdata_par} shows the results of the applied distance quality criteria.

        The first number indicates the \citet{Lindegren2018} filter (1 means that it is validated, and 0 that it is not validated), and the second number indicates our own condition (1 means that the {\em Gaia} distance is adopted, and 0 that the photometric distance is adopted). $FLAG_ {\rm D}=1 1$ means that both quality criteria are satisfied and we adopted {\em Gaia} parallaxes.

        \begin{figure}[htbp]
                \begin{center}
                        \includegraphics[width=0.5\textwidth,height=0.3\textheight,angle=0]{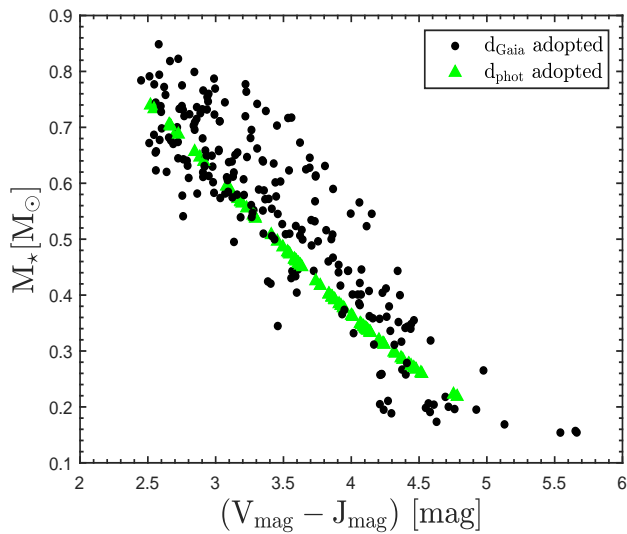}
                        \centering
                        \caption{Stellar masses as a function of $\rm V_{mag}-J_{mag}$ for all 302 stars analyzed in this work. We distinguish stars for which we adopted {\em Gaia} distances (black filled circles) and those for which we adopted photometric distances (green filled triangles) following the criteria described in Sect.~\ref{stellar_par}.}
                        \label{Mass_VJ}
                \end{center}
        \end{figure}
        
         We used the empirical relations from \citet{Mann2015} and \citet{Mann_2016} to calculate stellar parameters. In particular, we obtained stellar masses ($M_{\star}$) and radii ($R_{\star}$) from $M_{\rm K_{s}}$, the effective temperature ($T_{\rm eff}$) from $V-J$ and $J-H$,  the bolometric correction ($BC_{\rm K_{s}}$) from $V-J$, and the bolometric lumonosity $L_{\rm bol}$ from $BC_{\rm K_{s}}$. We list the stellar parameters fot the 14 new observations in Table~\ref{Par} and those for the literature sample in Table~\ref{ALLdata_par}.
         In Fig. \ref{Mass_VJ} we show the relation between $M_{\star}$ and $V-J$ we found for the full sample. We note that according to our $V-J$ versus SpT calibration \citep{Stelzer2016}, stars with $V-J<3$\, mag are K-type stars. For the sake of simplicity, we do not distinguish these objects, but we recall that the full sample comprises $\approx$15\% of K-type stars.

\begin{table*}
        \begin{center}
                \begin{threeparttable}[b]
                \caption{Stellar parameters of the 14 M dwarfs with new X-ray and rotation data.} 
                \small\addtolength{\tabcolsep}{-4pt}
                        \begin{tabular}{ccccccccccc}
                                \midrule[0.5mm]
                                \multicolumn{1}{c}{\textbf{K2 EPIC ID}} 
                                &\multicolumn{1}{c}{\textbf{$M_{\rm K_{s}}$}}  
                                &\multicolumn{1}{c}{\textbf{$M_{\star}$}}  
                                &\multicolumn{1}{c}{\textbf{$R_{\star}$}} 
                                &\multicolumn{1}{c}{\textbf{$M_{\rm bol}$}}
                                &\multicolumn{1}{c}{\textbf{$\logten\left(\frac{L_{\rm bol}}{L_{\odot}}\right)$}}
                                &\multicolumn{1}{c}{\textbf{$T_{\rm eff}$}} 
                                &\multicolumn{1}{c}{\textbf{$V-J$}}
                                &\multicolumn{1}{c}{\textbf{$P_{\rm rot}$}}
                                &\multicolumn{1}{c}{\textbf{$D$}}
                                &\multicolumn{1}{c}{\textbf{$FLAG_{\rm D}$}\tnote{1}} \\
                                
                                \multicolumn{1}{c}{} 
                                &\multicolumn{1}{c}{\textbf{[mag]}}  
                                &\multicolumn{1}{c}{\textbf{[$\rm M_{\odot}$]}}  
                                &\multicolumn{1}{c}{\textbf{[$\rm R_{\odot}$]}} 
                                &\multicolumn{1}{c}{\textbf{[mag]}}  
                                &\multicolumn{1}{c}{}
                                &\multicolumn{1}{c}{\textbf{[K]}} 
                                &\multicolumn{1}{c}{\textbf{[mag]}}
                                &\multicolumn{1}{c}{\textbf{[d]}}
                                &\multicolumn{1}{c}{\textbf{[pc]}}
                                &\multicolumn{1}{c}{}\\
                                \midrule[0.5mm]
                                
                                        201718613&7.01$\pm$0.06&0.31$\pm$0.01&0.31$\pm$0.01&9.75$\pm$0.07&-2.00$\pm$0.03&3260$\pm$89\phantom{0}&4.23&78.70&12.72$\pm$0.42&0 0\\
                                        212560714&5.54$\pm$0.03&0.54$\pm$0.01&0.52$\pm$0.02&8.04$\pm$0.05&-1.31$\pm$0.02&3906$\pm$93\phantom{0}&2.76&27.57&36.00$\pm$0.05&1 1\\
                                        214787262&6.63$\pm$0.04&0.37$\pm$0.01&0.36$\pm$0.01&9.33$\pm$0.05&-1.83$\pm$0.02&3382$\pm$80\phantom{0}&3.93&43.70&27.82$\pm$0.07&1 1\\
                                        201659529&6.58$\pm$0.03&0.37$\pm$0.01&0.36$\pm$0.01&9.28$\pm$0.05&-1.81$\pm$0.02&3355$\pm$80\phantom{0}&3.95&44.24&23.39$\pm$0.05&1 1\\
                                        202059222&6.76$\pm$0.05&0.35$\pm$0.01&0.34$\pm$0.01&9.48$\pm$0.06&-1.89$\pm$0.02&3308$\pm$80\phantom{0}&4.08&71.95&26.27$\pm$0.61&0 0\\
                                        202059188&6.75$\pm$0.04&0.35$\pm$0.01&0.34$\pm$0.01&9.52$\pm$0.05&-1.91$\pm$0.02&3179$\pm$81\phantom{0}&4.44&\phantom{0}0.69&28.74$\pm$0.07&1 1\\
                                        202059195&6.37$\pm$0.04&0.40$\pm$0.01&0.39$\pm$0.01&9.12$\pm$0.05&-1.74$\pm$0.02&3311$\pm$82\phantom{0}&4.23&42.79&34.61$\pm$0.10&1 1\\
                                        202059210&4.47$\pm$0.06&0.72$\pm$0.02&0.71$\pm$0.02&6.97$\pm$0.07&-0.89$\pm$0.03&3926$\pm$84\phantom{0}&2.81&17.40&54.37$\pm$0.24&1 1\\
                                        201364753&5.30$\pm$0.03&0.58$\pm$0.01&0.56$\pm$0.02&7.82$\pm$0.05&-1.23$\pm$0.02&3854$\pm$87\phantom{0}&2.86&\phantom{0}9.19&40.96$\pm$0.09&1 1\\
                                        202059198&5.81$\pm$0.02&0.50$\pm$0.01&0.47$\pm$0.01&8.39$\pm$0.04&-1.45$\pm$0.02&3727$\pm$99\phantom{0}&3.14&26.93&23.30$\pm$0.03&1 1\\
                                        210579749&5.51$\pm$0.02&0.55$\pm$0.01&0.52$\pm$0.02&8.10$\pm$0.04&-1.34$\pm$0.02&3643$\pm$83\phantom{0}&3.27&11.16&17.24$\pm$0.01&1 1\\     
                                        214269391&4.95$\pm$0.02&0.64$\pm$0.01&0.62$\pm$0.02&7.44$\pm$0.05&-1.08$\pm$0.02&3900$\pm$101&2.76&19.56&17.66$\pm$0.01&1 1\\
                                        203869467&4.68$\pm$0.04&0.69$\pm$0.01&0.67$\pm$0.02&7.17$\pm$0.06&-0.97$\pm$0.02&3951$\pm$84\phantom{0}&2.73&47.58&39.07$\pm$0.94&0 0\\
                                        201717791&4.59$\pm$0.04&0.7$\pm$0.01&0.69$\pm$0.02&7.10$\pm$0.05&-0.94$\pm$0.02&3873$\pm$90\phantom{0}&2.84&14.40&46.45$\pm$0.13&1 1\\
                                        
                                        \bottomrule[0.5mm]
                                \end{tabular}
                        \label{Par}
                        \begin{tablenotes}
                        \item[1] $FLAG_{\rm D}$ is the quality criteria we used to select the distance. The first column represents the quality flag of {\em Gaia} parallaxes from \citet{Lindegren2018} (1 means that it is validated, and 0 that it is  not validated), the second column shows our criteria for the comparison between {\em Gaia} and photometric distances, explained in Sect.~\ref{stellar_par} (1 means that the {\em Gaia} distance is adopted, and 0 that the photometric distance is adopted). We used {\em Gaia} parallaxes if $FLAG_{\rm D} = 1 1$.
                        \end{tablenotes}
                \end{threeparttable}
        \end{center}
\end{table*}

        Because we updated the distances for the literature samples, we had to recalculate the X-ray luminosity ($L_{\rm x}$). In order to have a uniform data sample, we computed the $L_{\rm x}$ adopting the {\em ROSAT} energy band ($0.1-2.4$\,keV) used in \citet{Wright2011}, \citet{Wright2016}, and \citet{Wright2018} for the full sample. Of the objects in \citet{Wright2011} we took only the field stars, and we scaled the published $L_{\rm x}$ values with $\left(\frac{d_{\rm new}}{d_{\rm Wr+11}}\right)^2$, where $d_{\rm new}$ is our new distance from Table~\ref{ALLdata_par} and $d_{\rm Wr+11}$ is the distance used by \citet{Wright2011}. For the stars from \citet{Wright2016} and \citet{Wright2018}, we calculated $L_{\rm x}$ from the fluxes listed in \citet{Wright2018} with our new distances. Because \citet{Gonzalez-Alvarez2019} listed $L_{\rm x}$ for $0.1-2.0$\,keV, in order to obtain the X-ray luminosity in the $0.1-2.4$\,keV band, we returned to the {\em ROSAT} catalogs. In particular, we extracted the count rates from the bright \citep[BSC:][]{Voges1999} and faint \citep[FSC:][]{Voges2000} source catalogs and the second {\em ROSAT} All-Sky Survey Point Source Catalog \citep[2RXS:][]{Boller2016}. These were converted into X-ray flux using the conversion factor ($CF=5.771\cdot 10^{-12}\hspace{1mm} \rm erg/cm^{2}/cts$) obtained with the count-rate simulator WebPIMMS\footnote{Count-rate simulator PIMMS:\\
        \url{https://cxc.harvard.edu/toolkit/pimms.jsp}} for a 1T-APEC model with $\rm kT=0.5$~keV and $\rm N_{H} = 10^{19} cm^{-2}$. The temperature value is derived from computingf the mean coronal temperature for the stars from our new {\em XMM-Newton} and {\em Chandra} sample that have enough counts for the spectrum to be extracted (see Sect.~\ref{spectra}), for which we find an average of $0.51\pm0.03$~keV.

        The uncertainties of the X-ray luminosities were calculated with error propagation,  using the variance formula for the uncertainties of the X-ray fluxes and distances. \citet{Wright2011} provided no uncertainties on the X-ray measurements, therefore we applied the mean percentage value of the X-ray flux error measured for the other samples, which is $\approx$15\%.
        
        We list in Table~\ref{ALLdata_par} our updated results for the X-ray luminosities of the 288 stars from the literature. We also provide (in Col.7) the rotation period adopted from these previous studies. In particular, \citet{Wright2011} selected only photometric  $P_{\rm rot}$ from the literature, \citet{Wright2016} and \citet{Wright2018} used rotation measurements from the MEarth Project, which means that they were mostly in the slow-rotator regime. \citet{Stelzer2016} have determined the $P_{\rm rot}$ for the LG11 stars in the {\em K2} field of campaigns C0 toC4, and \citet{Gonzalez-Alvarez2019} analyzed time-series of high-resolution spectroscopy taking $P_{\rm rot}$ from activity indicators, that is, the Ca{\footnotesize II} H\&K and H{\footnotesize $\alpha$} spectral lines.
        
\begin{figure*}
        \begin{multicols}{2}
                \includegraphics[width=0.45\textwidth]{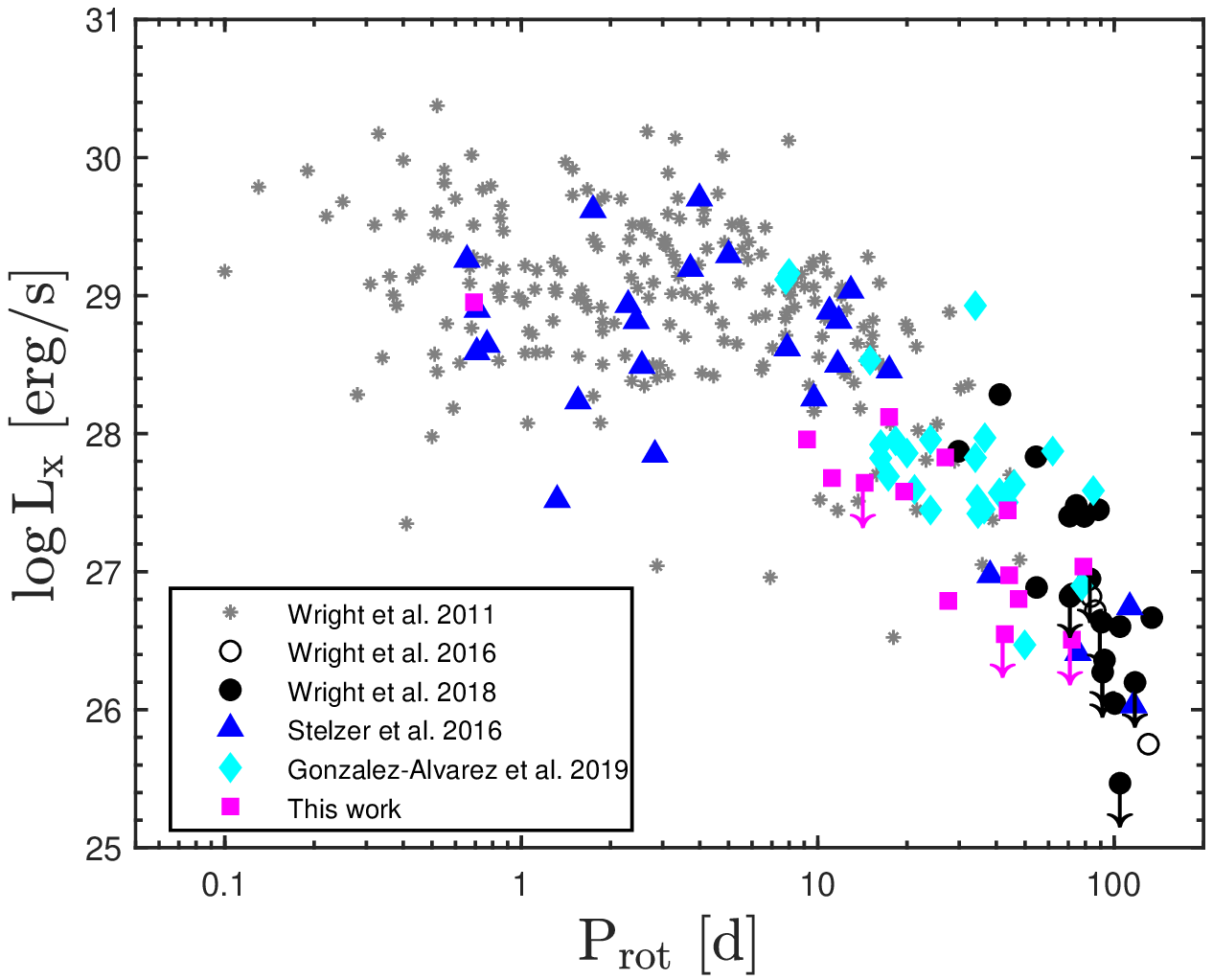}\par 
                \includegraphics[width=0.45\textwidth]{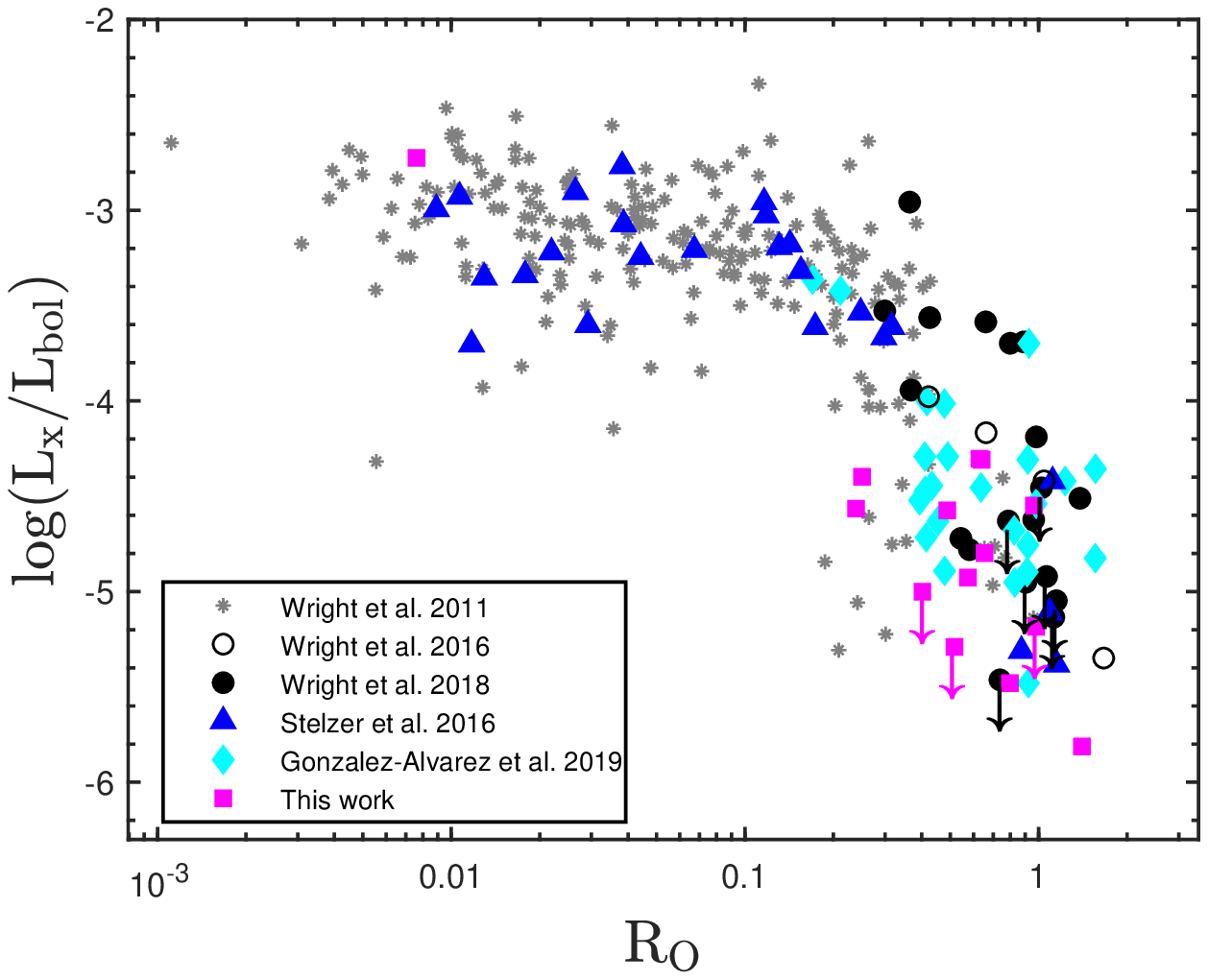}\par 
        \end{multicols}
        \caption{\textbf{Left}: Relation of activity to rotation  in $L_{x}-P_{\rm rot}$ space. The color scale is based on the origin of the sample, as explained in the inset. \textbf{Right}: Same as the left panel, but for $L_{x}/L_{\rm bol}$ vs. Rossby number.}
        \label{LxLbolRo_Ref}
\end{figure*}

        To illustrate the different samples, we show in Fig.~\ref{LxLbolRo_Ref} the updated relation of X-ray activity  to rotation by combining all previous literature samples with our own data, listed in Tables~\ref{Par} and \ref{ALLdata_par}.
For this plot the convective turnover times, $\tau_{\rm conv}$, were calculated using the empirical calibration by \citet{Wright2018}.

\section{X-ray data analysis}
\label{data_analysis}
        As explained above, we worked in the {\em ROSAT} energy band (0.1-2.4~keV) for consistency with most previous works. The results from the analysis of X-ray data for the new sample of 14 stars obtained with {\em Chandra} and {\em XMM-Newton} are listed in Table~\ref{XRay_Results}. 
        In the following we describe the analysis of these observations.
        \begin{table*}
                 \begin{center}
                        \caption{X-ray journal of observations together with the results from the analysis explained in Sects.~\ref{data_analysis} and \ref{LxLbolRo_space}. The $R_{\rm O,C\&S}$ and $R_{\rm O,B}$ columns are the $R_{\rm O}$ numbers from the normalized $\tau_{\rm conv}$ relations by \citet{Cranmer_2011} and \citet{Brun2017} for stars with $T_{\rm eff} > 3300$~K (see Sect.~\ref{LxLbolRo_space} for more details). The last column ($R_{\rm O,W}$) shows the $R_{\rm O}$ number by \citet{Wright2018}.}  
                        \small\addtolength{\tabcolsep}{-4pt}
                                \begin{tabular}{ccccccccccc}
                                \midrule[0.5mm]                    
                                \multicolumn{1}{c}{\textbf{K2 EPIC ID}} 
                                &\multicolumn{1}{c}{\textbf{Mission}}  
                                &\multicolumn{1}{c}{\textbf{Obs. ID}} 
                                &\multicolumn{1}{c}{\textbf{Obs. Date}}  
                                &\multicolumn{1}{c}{\textbf{Exp. Time}} 
                                &\multicolumn{1}{c}{\textbf{Rate}}
                                &\multicolumn{1}{c}{\textbf{$\log(L_{x}$)}}
                                &\multicolumn{1}{c}{\textbf{$\log\left(\frac{L_{x}}{L_{bol}}\right)$}}
                                &\multicolumn{1}{c}{\textbf{$R_{\rm O,C\&S}$}}
                                &\multicolumn{1}{c}{\textbf{$R_{\rm O,B}$}}
                                &\multicolumn{1}{c}{\textbf{$R_{\rm O,W}$}}\\

                                \multicolumn{1}{c}{} 
                                &\multicolumn{1}{c}{}  
                                &\multicolumn{1}{c}{} 
                                &\multicolumn{1}{c}{}  
                                &\multicolumn{1}{c}{\textbf{[ks]}} 
                                &\multicolumn{1}{c}{\textbf{[$\cdot 10^{-3}$counts/s]}}
                                &\multicolumn{1}{c}{\textbf{[erg /s]}}
                                &\multicolumn{1}{c}{}
                                &\multicolumn{1}{c}{}
                                &\multicolumn{1}{c}{}
                                &\multicolumn{1}{c}{}\\
                                \midrule[0.5mm]
                                
                                201718613&XMM-Newton&0820460101&2018-06-11&18.0&47.20$\pm$2.20&27.03$\pm$0.03&-4.54$\pm$0.96&--&--&0.96\\
                                212560714&XMM-Newton&0820460201&2018-07-02&33.6&\phantom{0}3.33$\pm$0.57&26.78$\pm$0.07&-5.48$\pm$0.09&0.29&0.37&0.80\\
                                214787262&XMM-Newton&0820460301&2019-03-31&45.5&25.15$\pm$1.27&27.44$\pm$0.02&-4.30$\pm$0.10&0.35&0.30&0.64\\
                                201659529&XMM-Newton&0843430401&2019-07-14&23.1&12.07$\pm$1.05&26.97$\pm$0.04&-4.79$\pm$0.09&0.35&0.31&0.65\\
                                202059222&Chandra&17724&2015-12-07&14.7&<\phantom{0}0.24&<26.50&<-5.20&0.56&0.41&0.98\\
                                202059188&Chandra&17725&2016-01-12&14.9&56.20$\pm$3.20&28.95$\pm$0.02&-2.70$\pm$0.10&--&--&0.01\\
                                202059195&Chandra&17726&2015-12-02&15.0&<\phantom{0}0.15&<26.54&<-5.30&0.33&0.34&0.52\\
                                202059210&Chandra&17727&2015-11-27&13.9&\phantom{0}2.32$\pm$0.81&28.12$\pm$0.15&-4.60$\pm$0.10&0.19&0.40&0.49\\
                                201364753&Chandra&17728&2016-03-26&\phantom{0}8.8&\phantom{0}0.52$\pm$0.43&27.95$\pm$0.11&-4.40$\pm$0.10&0.09&0.17&0.25\\
                                " " &Chandra&18805&2016-03-31&\phantom{0}5.8&\phantom{0}2.36$\pm$1.10&" "& " "& " "& " "& " "\\
                                202059198&Chandra&17729&2016-02-05&14.8&\phantom{0}6.42$\pm$0.36&27.82$\pm$0.02&-4.30$\pm$0.10&0.26&0.32&0.63\\
                                210579749&Chandra&21157&2018-10-22&\phantom{0}9.4&\phantom{0}3.24$\pm$0.97&27.67$\pm$0.13&-4.60$\pm$0.10&0.10&0.18&0.24\\
                                214269391&Chandra&21158&2018-11-05&\phantom{0}9.9&\phantom{0}2.46$\pm$0.84&27.58$\pm$0.14&-4.90$\pm$0.10&0.21&0.37&0.57\\
                                203869467&Chandra&21159&2019-01-20&27.7&\phantom{0}0.08$\pm$0.25&26.80$\pm$0.02&-5.80$\pm$0.01&0.51&0.84&1.40\\
                                201717791&Chandra&21160&2018-11-06&15.7&<\phantom{0}0.41&<27.64&<-5.00&0.15&0.33&0.40\\
                        \bottomrule[0.5mm]
                        \end{tabular}
                \label{XRay_Results}
        \end{center}
\end{table*}
\subsection{XMM-Newton: EPIC}
\label{EPIC}
        Four of the 14 new observations were obtained with {\em XMM-Newton}. We analyzed these 4 observations with the {\em XMM-Newton} Science Analysis System (SAS)\footnote{SAS Data Analysis Threads: \\
        \url{https://www.cosmos.esa.int/web/xmm-newton/sas-threads}} 17.0 pipeline. After data extraction, we filtered the event lists of EPIC/pn and EPIC/MOS. 
        We extracted the light curve for the whole detector, and then we determined the good time intervals using the count rate $\leq0.4$~cts/s and count rate $\leq0.35$~cts/s as threshold for EPIC/pn and EPIC/MOS, respectively. We further filtered our data for pixel pattern (PATTERN=0), event energies greater than 0.15~keV, and quality flag (FLAG=0). 
        We performed source detection using the SAS pipeline $edetect\_chain$ in the {\em ROSAT} energy band (0.15-2.4~keV) simultaneously for EPIC/pn and EPIC/MOS.
        Our lower energy threshold is slightly different from that of the {\em ROSAT} band. This is based on the fact that the recommended low-energy cutoff for {\em XMM-Newton} is at 0.15~keV. However, we quantified how much the count rate would differ if we included the counts between $0.10-0.15$\,keV, and it would be only $\approx 2\%$ greater.
        The extraction of spectra and light curves was performed considering a source region of 40$''$ centered on the source position with an adjacent source-free circular background region three times greater. We created the response matrix and ancillary response for the spectral analysis with the SAS tools \textit{RMFGEN} and \textit{ARFGEN}, and we rebinned the spectrum in order to have at least five counts for each background-subtracted spectral channel. 
\subsection{Chandra}
        {\em \textup{The} Chandra} data analysis was carried out with the CIAO package\footnote{The CIAO package is developed by the {\em Chandra} X-Ray Center for analyzing data from the {\em Chandra} X-ray Telescope, it can be downloaded from \url{http://cxc.harvard.edu/ciao/}}. We started our analysis with the new pipeline \textit{chandra\_repro}, which automatically reprocesses the event list by reading data from the standard data distribution and creating a new bad pixel file and a new level 2 event file. After this, we created an exposure-corrected image for CCD\_ID = 7 of our ACIS-S observations in the {\em ROSAT} energy band (0.1 to 2.4 keV), and we determined the point spread function (PSF) map of the image with \textit{mkpsfmap}, choosing 100\% of the enclosed counts fraction (ecf=1.0).
        At this point, we proceeded with the source detection with the \textit{wavdetect} algorithm, which takes into account the PSF map, the exposure time, and a significance detection threshold, which we set to $\sigma=10^{-5}$. This value is needed to identify a pixel as belonging to a source. Three of the ten stars observed with {\em Chandra} are undetected, and we calculated the flux upper limits.
        We calculated the count rates using \textit{srcflux} for detected and undetected stars, giving the positions, the source, and background regions and the {\em ROSAT} energy band as inputs. In particular, we took the circular source region centered on our sources and a circular region for the background 10-15 times greater than the source regions. 
        For undetected sources, \textit{srcflux} computes the upper limit count rate using the Bayesian posterior probability distribution function, without assuming prior information for the intensities in the source and background apertures.
        
\subsection{X-ray spectra}
\label{spectra}
\begin{figure}[htbp]
                \includegraphics[width=0.5\textwidth]{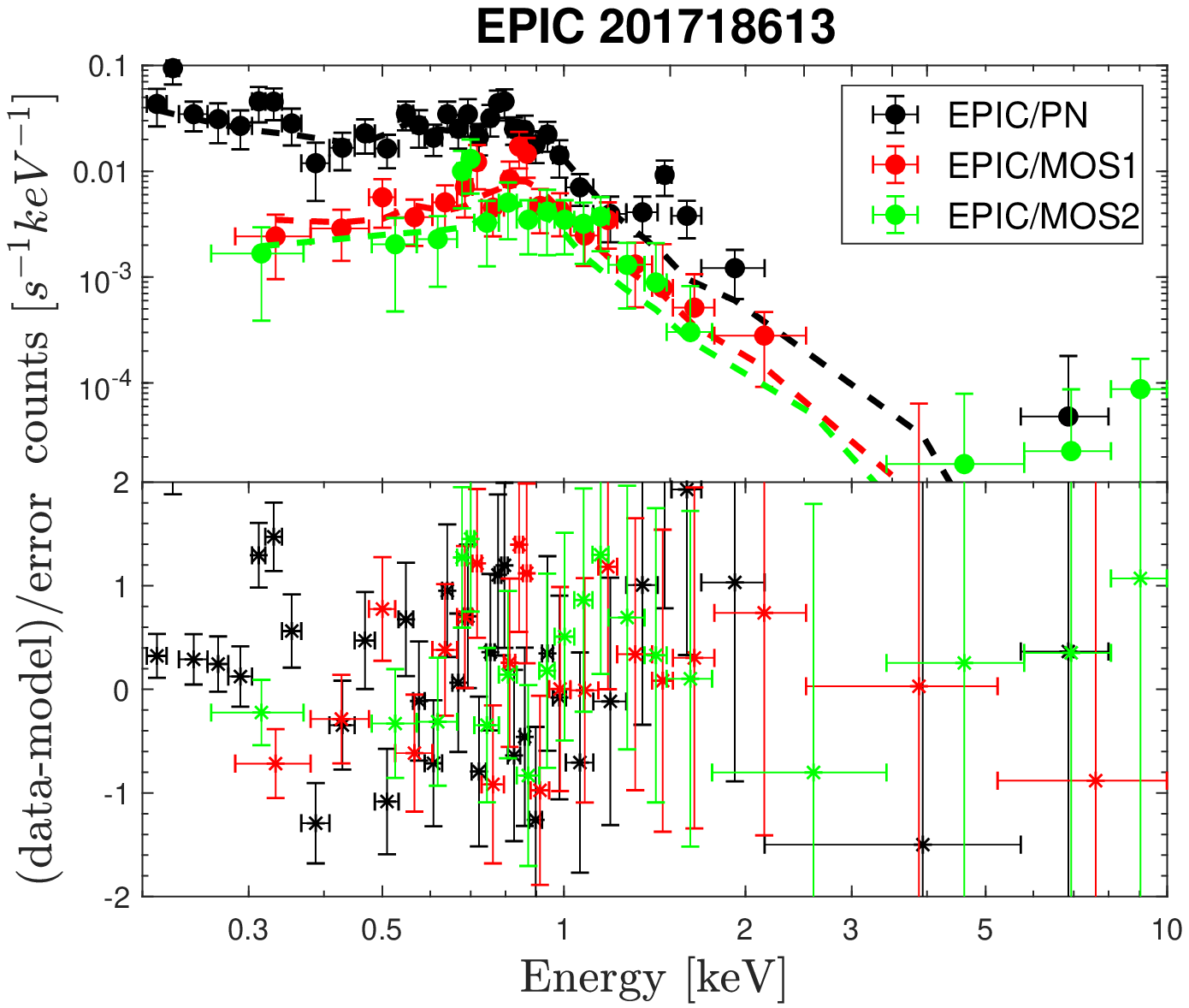}\\
                \includegraphics[width=0.5\textwidth]{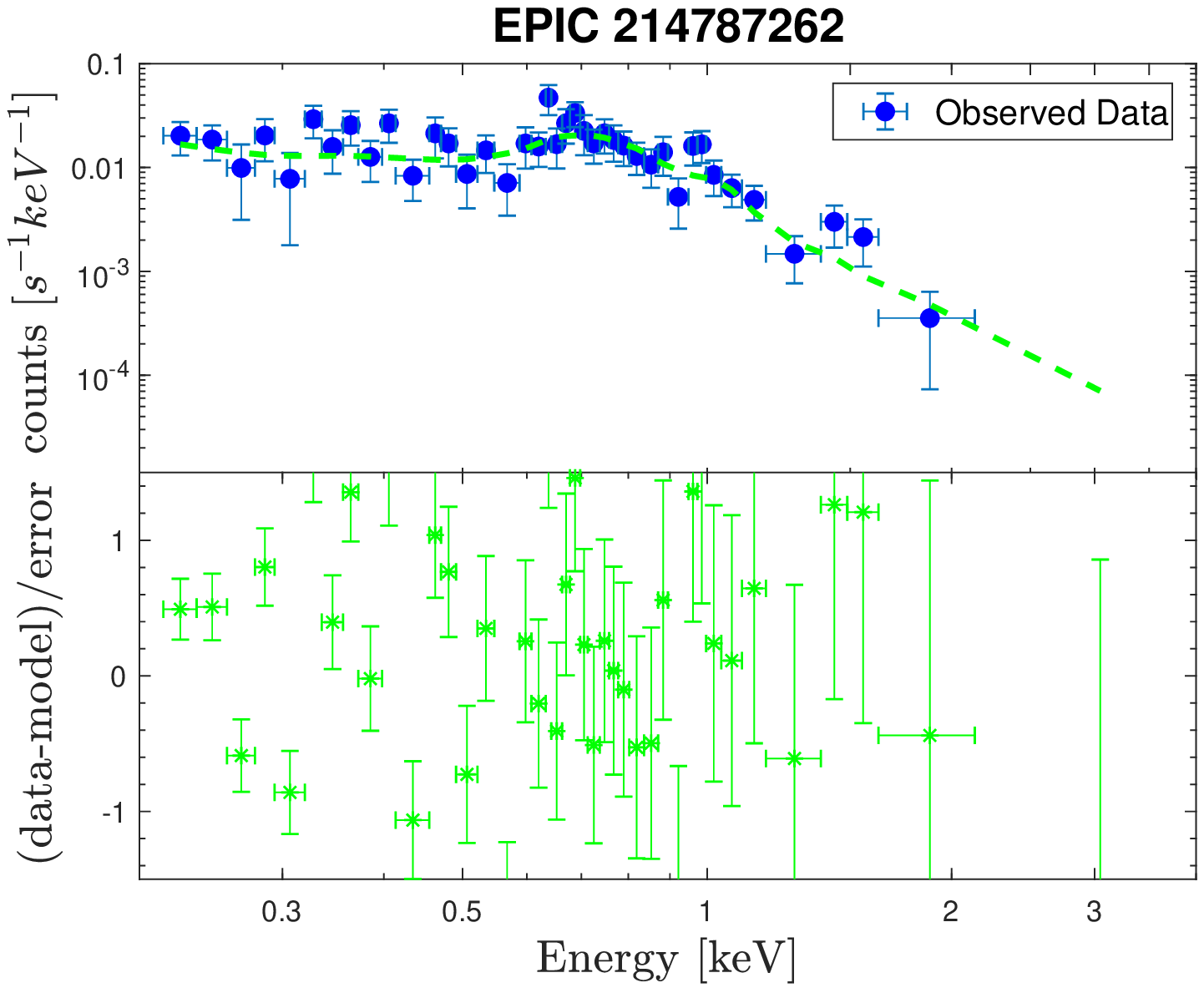}
                \caption{EPIC X-ray spectra together with the best-fitting thermal APEC model (dashed line) for the two {\em XMM-Newton} observations with sufficient counts for spectral analysis. In particular, a simultaneously multifit for EPIC/pn and EPIC and MOS spectra of EPIC\,201718613 and a single fit, again with two temperatures, for EPIC/pn data for EPIC\,214787262 are shown.}
                \label{spec}
        \end{figure}

        Spectral analysis was performed with XSPEC\footnote{XSPEC NASA's HEASARC Software:\\ \url{https://heasarc.gsfc.nasa.gov/xanadu/xspec/}} version 12.10, fitting the two extracted spectra with more than 350 counts, using two isothermal \rm{APEC} models. Each APEC model has three parameters: the plasma temperature ($kT$), the global abundance ($Z$), and the emission measure ($EM$). We fixed Z at $0.3~Z_{\odot}$, the typical global abundance for late-type stars, and we left $kT$ and $EM$ free to vary. In particular, we performed a multi-fitting procedure for EPIC~201718613 by simultaneously fitting the spectra from the three instruments on board {\em XMM-Newton}. On the other hand, for EPIC~214787262, we fit only the EPIC/pn spectrum because  EPIC and MOS have not enough counts to extract the spectra.
        The parameters of the best-fitting model are listed in Table~\ref{xspec_Xmm_output} and the spectra are shown in Fig.~\ref{spec}.
        \begin{table*}
                \caption{X-ray spectral parameters with 1~$\sigma$ uncertainties computed with the error pipeline provided in the XSPEC software package.}            
                \begin{center}
                        \begin{tabular}{cccccccc}
                                \midrule[0.5mm]                    
                                \multicolumn{1}{c}{\textbf{K2 EPIC ID}} 
                                &\multicolumn{1}{c}{\textbf{$kT_{1}$}} 
                                &\multicolumn{1}{c}{\textbf{$\log\left(EM_{1}\right)$}}  
                                &\multicolumn{1}{c}{\textbf{$kT_{2}$}} 
                                &\multicolumn{1}{c}{\textbf{$\log\left(EM_{2}\right)$}}
                                &\multicolumn{1}{c}{\textbf{$\chi^{2}$}}
                                &\multicolumn{1}{c}{\textbf{d.o.f.}}
                                &\multicolumn{1}{c}{\textbf{$T_{\rm mean}$}}\\  

                                \multicolumn{1}{c}{} 
                                &\multicolumn{1}{c}{\textbf{[keV]}}  
                                &\multicolumn{1}{c}{\textbf{[cm$^{-3}$]}}  
                                &\multicolumn{1}{c}{\textbf{[keV]}} 
                                &\multicolumn{1}{c}{\textbf{[cm$^{-3}$]}}  
                                &\multicolumn{1}{c}{}
                                &\multicolumn{1}{c}{} 
                                &\multicolumn{1}{c}{\textbf{[keV]}}\\
                                \midrule[0.5mm]
                                201718613 & 0.17$\pm$0.03 & 48.54$\pm$0.07 & 0.73$\pm$0.05 & 48.49$\pm$0.07&0.8&69&0.44$\pm$0.03\\
                                214787262 & 0.31$\pm$0.02 & 50.17$\pm$0.05 & 1.29$\pm$0.21 & 49.77$\pm$0.12&1.1&37&0.59$\pm$0.06\\
                                \bottomrule[0.5mm]
                        \end{tabular}
                        \label{xspec_Xmm_output}
                \end{center}
        \end{table*}
        The emission measure is computed in logarithmic scale, and it is the square of the number density of free electrons integrated over the volume of the plasma. 
        With the EM, we computed the mean coronal temperature ($T_{\rm mean}$). In particular, $T_{\rm mean}$ is defined as 
        \begin{equation}
        T_{\rm mean} = \frac{\sum\left(\rm EM_{\rm n}\cdot \rm T_{\rm n}\right)}{\sum\left(\rm EM_{\rm n}\right)},\end{equation} 
        where $T_{\rm n}$ and $EM_{\rm n}$ are the n-temperatures and n-EM of the fitted model.
        From the two $T_{\rm mean}$ (see Table~\ref{xspec_Xmm_output}), we found the average $\rm kT=0.51\pm0.03$~keV that we used together with $\rm N_{H} = 10^{19} cm^{-2}$ to compute the conversion factors (CF) with WebPIMMS needed to determine the X-ray fluxes. In particular, we calculated the flux in the {\em ROSAT} energy band (0.1-2.4~keV) for the full sample, but for stars observed with {\em XMM-Newton,} the fluxes were extracted in the readapted energy band ($0.15-2.4$\,keV), as explained in Sect.~\ref{EPIC}. 
        In particular, for {\em Chandra} cycle 17, we found $\rm CF = 1.61\cdot 10^{-11} \rm erg/cm^{2}/cts$, for {\em Chandra} cycle 20, we found $\rm CF = 4.14\cdot 10^{-11} \rm erg/cm^{2}/cts$, and for {\em XMM-Newton,} we found $\rm CF = 1.19\cdot 10^{-12} \rm erg/cm^{2}/cts$.

\section{K2 analysis}
\label{K2_an}
Our 14 targets were selected because they have photometric monitoring observations by the {\em K2} mission. {\em K2} observed in two cadence modes, long cadence ($\sim$30\,min data point cadence) and short cadence ($\sim$1\,min data point cadence). While all 14 targets have light curves obtained in long-cadence mode, 3 targets were also observed with the $\sim$1\,min data point cadence.
We downloaded the fully reduced and corrected long-cadence light curves provided by \citet{2014PASP..126..948V} from the website of A.~Vanderburg\footnote{\url{https://www.cfa.harvard.edu/~avanderb/k2.html}}. The rotation periods were measured using standard time-series analysis techniques, that is, the generalized Lomb-Scargle periodogram \citep[\begin{small}GLS\end{small};][]{2009A&A...496..577Z}, the autocorrelation function (ACF), and the fitting of the light curves with a sine function. While GLS and ACF are limited to periods shorter than the campaign duration of 70-80\,d, the sine fitting allows us to constrain rotation periods even if they exceed the {\em K2} monitoring time baseline, as is the case for EPIC\,202059222, for example. For each target we obtained three estimates for the rotation period. Through by-eye inspection of the phase-folded light curves from each method, we selected the best-fitting period. When several methods yielded equally good periods, we adopted the average rotation period as the final value.
A detailed description of our procedure for measuring rotation periods can be found in \citet{Raetz2020}. Our final adopted values of the rotation periods are summarized in Table~\ref{Par}. The periods were found to agree within $<$5\% with the values published by \citet{Stelzer2016}, Raetz et al. (2020a, AN submitted), and \citet{Raetz2020}..

\section{Relation of activity, rotation, and age }
\label{results}
In this section we discuss the relation of X-ray activity, rotation, and age based on the full sample as is defined in Sect.~\ref{stellar_par}. We use the result combined with angular momentum evolution models by \citet{Matt_2015} to construct the time-evolution of the X-ray luminosity of M dwarfs.
As we explained in Sect.~\ref{intro}, previous studies showed two different regimes of the rotation-activity relation, the saturated regime for fast-rotating stars with $P_{\rm rot} \leq P_{\rm rot_{sat}}$ and the unsaturated regime for slowly rotating stars with $P_{\rm rot} > P_{\rm rot_{sat}}$. The convective turnover time rescales the sample by decreasing the horizontal spread in the unsaturated regime and shifting the break point between the saturated and unsaturated regime; normalizing the X-ray luminosity by the stellar bolometric luminosity decreases the vertical spread in both regimes, making 
the distinction of the two regimes more pronounced in the $L_{x}/L_{\rm bol}-R_{O}$ space.

 In Fig.~\ref{Act_Prot} we show the full sample, plotted with a color-scale representing the stellar mass. Arrows denote upper limits. Three of these undetected sources come from our new X-ray data and seven are from \citet{Wright2018} (see Tables~\ref{XRay_Results} and \ref{ALLdata_par} for more details).
The best parameters for characterizing the relation between activity and rotation have long been debated. Here we study both the $L_{x}-P_{\rm rot}$ and $L_{x}/L_{\rm bol}-R_{O}$ relation in Sects.~\ref{LxProt_space} and \ref{LxLbolRo_space}, respectively.  
\subsection{X-ray luminosity versus rotation period}
\label{LxProt_space}
\begin{figure}[htbp]
        \begin{minipage}[c]{0.5\textwidth}
                \begin{center}
                        \includegraphics[width=\textwidth]{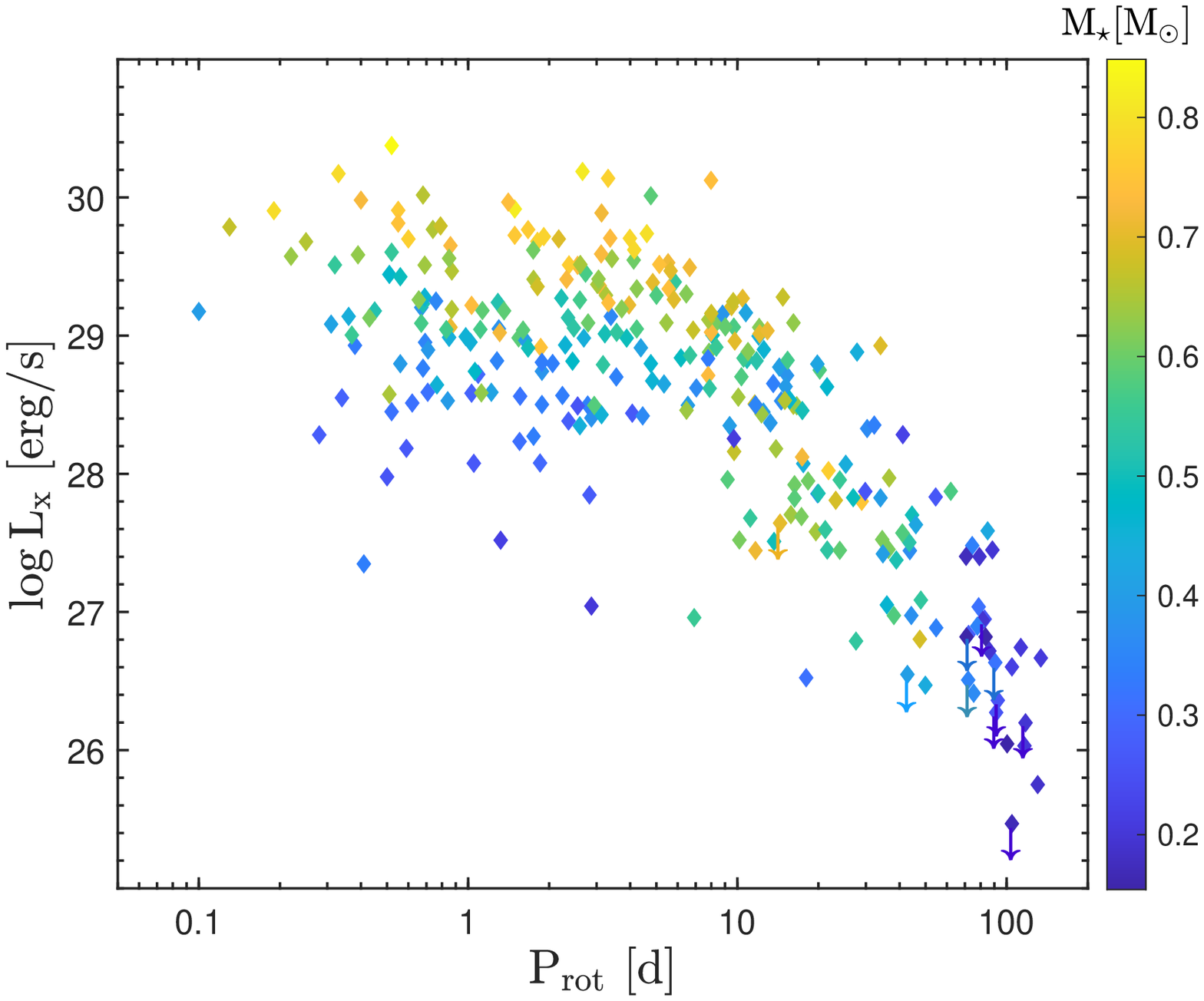}
                        \centering
                \end{center}
        \end{minipage}
        \begin{minipage}[c]{0.5\textwidth}
                \begin{center}
                        \includegraphics[width=\textwidth]{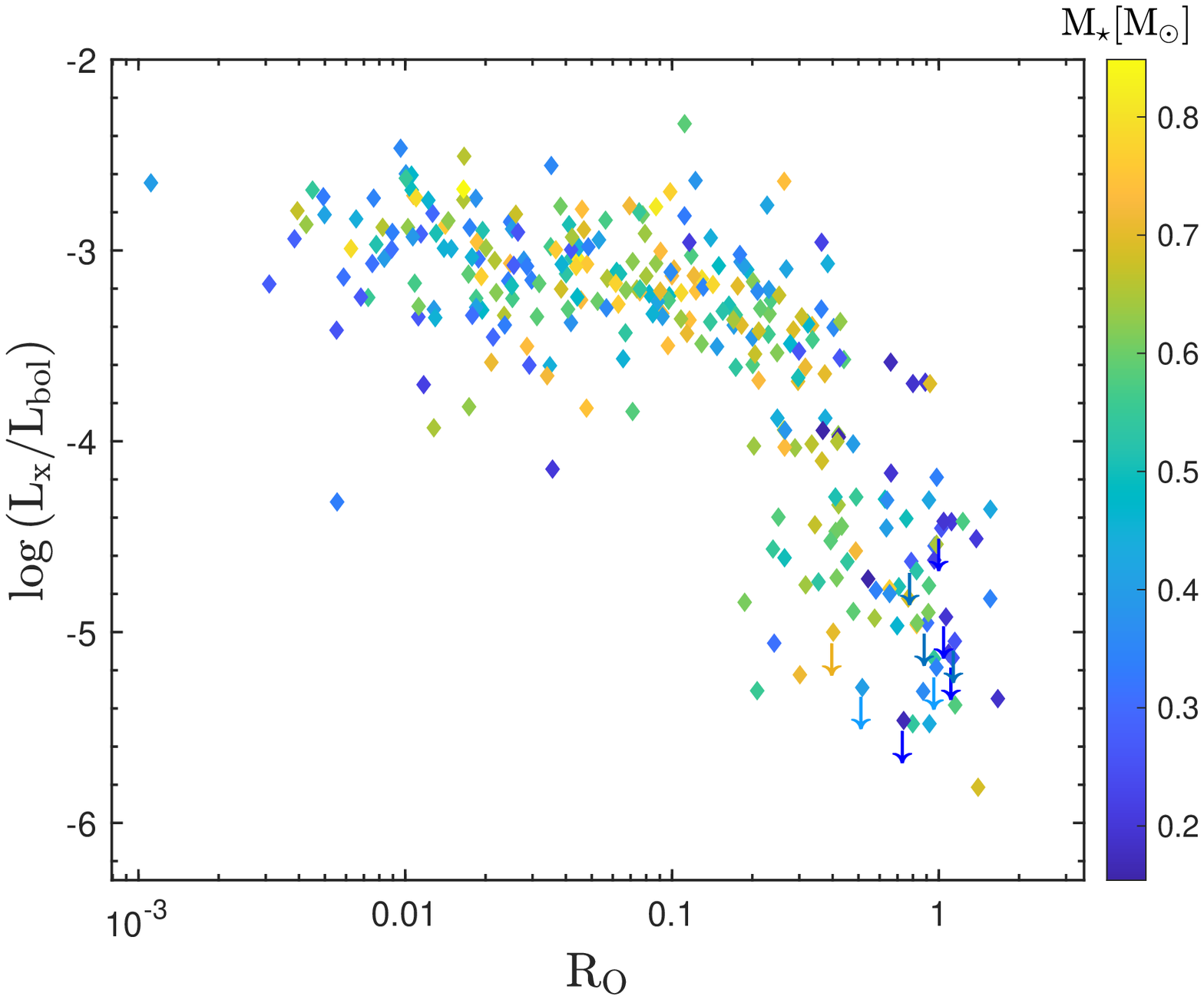}
                        \centering
                \end{center}
        \end{minipage}
        \caption{Relation of activity to rotation for all 302 stars we analyzed, displayed with a color code for the stellar mass. \textbf{Top}: Relation in X-ray luminosity vs. rotation period space. \textbf{Bottom}: Relation in terms of the ratio between X-ray and bolometric luminosities as a function of Rossby number.}
        \label{Act_Prot}
\end{figure}

The observed activity-rotation relation in $L_{x}-P_{\rm rot}$ space (Fig.~\ref{Act_Prot} top panel) shows a large vertical spread, amounting to $\approx$2~dex, in the saturated and unsaturated regime. Moreover, the X-ray activity in the saturated regime does not seem to show a constant maximum value, but the $L_{\rm x}$ level instead appears to decrease from a maximum at the shortest rotation periods, declining toward the breaking point into the unsaturated regime.
For this reason, our approach is based on a broken power-law fit for the two regimes. 

In particular, our fitting method requires three steps. We first use a Bayesian approach to infer the maximum likelihood parameters for a dual power law in the $L_{\rm x}$ versus $P_{\rm rot}$ space.  Our implementation of this dual power-law model is based on routines originally developed by \citet{Douglas2014} for use with the emcee Markov chain Monte Carlo (MCMC) package \citep{Foreman-Mackey2013} to infer the maximum likelihood parameters of the model.  In detail, the dual power-law fit is calculated as shown in Eq.~\ref{fit_kevin},

\begin{equation}
L_{\rm x} = 
\begin{cases} C_{sat\phantom{un}} P_{\rm rot}^{\beta_{sat}} & \mbox{if } P_{\rm rot}\leq  P_{\rm rot, sat}  \\  C_{unsat} P_{\rm rot}^{\beta_{unsat}}  & \mbox{if } P_{\rm rot} > P_{\rm rot, sat}
\end{cases}
\label{fit_kevin}
,\end{equation}
where $C_{n} = (L_{x_{n}}/P_{\rm rot}^{\beta_{n}})$, with n = (sat,unsat).

In our first iteration, likelihoods of each potential model are calculated using flat priors in each parameter ($2\, \rm d< P_{rot, sat}<50\, \rm d$; -4 $< \beta_{sat} <$ 2; -5 $< \beta_{unsat} <$ 1), and allowing for a nuisance parameter to account for underestimated (multiplicative) errors. We infer maximum likelihood parameters by comparing each potential model output to the subset of the full sample with reliable detections (i.e., excluding nondetections from this first iteration) using 256 walkers that each take 10,000 steps in their MCMC chain.  We discard the first half of each chain to allow the solutions to burn in, and measure the maximum likelihood values of each parameter as the median value of the remaining samples; we calculate 1$\sigma$ uncertainties as half the distance between the 16th$^{}$ and 84th$^{}$ percentiles of the resulting posterior distribution.  In practice, the latter nuisance parameter converged quite closely to 1, suggesting that the adopted uncertainties are appropriately close to their true values, therefore we do not report these values further.

In order to take the upper limits properly into account, in the next step we fit only the unsaturated regime, where all upper limits are located, using the Cenken method provided by the R-statistics package to calculate the Akritas-Theil-Sen \citep{Akritas1995} nonparametric slope to the full censored dataset. To define the onset of the unsaturated regime in terms of $P_{\rm rot}$, we used the result from the MCMC analysis in the previous step. To ensure that our measurement of the slope in the saturated regime was not unduly influenced by the omission of nondetections from the first MCMC fit, we then repeated the MCMC-based inference of the dual power-law fit, but forcing the slope in the unsaturated regime to remain within 0.02 of the value identified by the Cenken routine. 

\begin{figure}[htbp]
        \begin{center}
                \includegraphics[width=0.5\textwidth,height=0.3\textheight,angle=0]{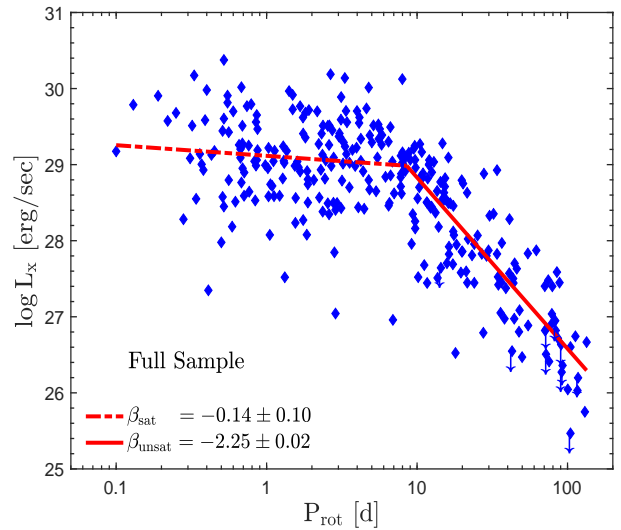}
                \caption{Two-component fit (see Eq.~\ref{fit_kevin}) to the activity-rotation relation for the full sample (see Sect.~\ref{LxProt_space} for the fitting procedure).}
                \label{LxvsProt_Fit_Full}
        \end{center}
\end{figure}

The results inferred from this three-step fitting process are shown in Fig.~\ref{LxvsProt_Fit_Full}, and tabulated for reference in Table~\ref{Fit_res}. As a result of this procedure, we found maximum likelihood parameters for a dual power-law fit to the full mass range of $\beta_{sat} = -0.14 \pm 0.10$, $\beta_{unsat} = -2.25 \pm 0.02,$ and $P_{\rm rot,sat} = 8.5 \pm 1.0$~d. We quantify for the first time that the X-ray luminosity in the saturated regime is not constant but shows a small negative slope, that is, a decrease in $L_{\rm x}$ for higher $P_{\rm rot}$. However, the uncertainties of $\beta_{\rm sat}$ indicate that this finding is tentative, with a significance at the $\sim1 \sigma$ level for the full global fit.

\begin{figure}
        \begin{center}
                \includegraphics[width=0.5\textwidth]{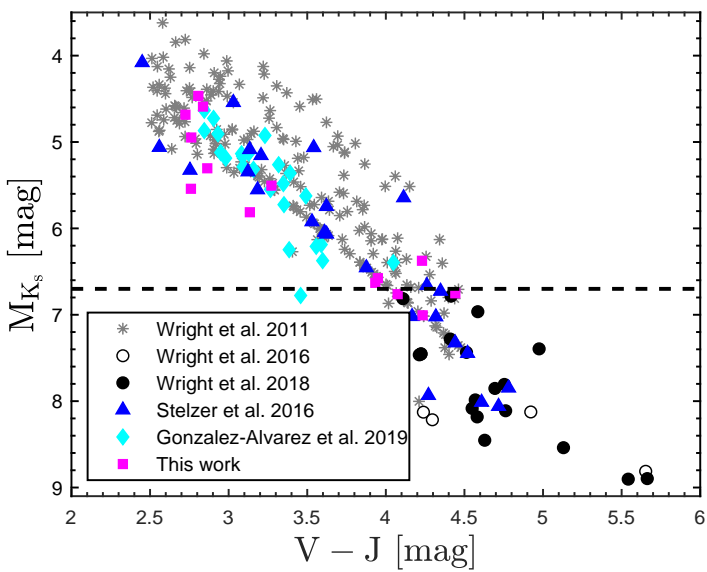}
                \caption{Color-magnitude diagram for the full sample. The dashed black line shows the transition to fully convective stars at $M_{\rm K_{s}}>6.7$~mag, according to \citet{Jao_2018}.}
                \label{Mks}
        \end{center}
\end{figure}    
\begin{figure*}[h!]
        \begin{multicols}{2}
                \includegraphics[width=\linewidth]{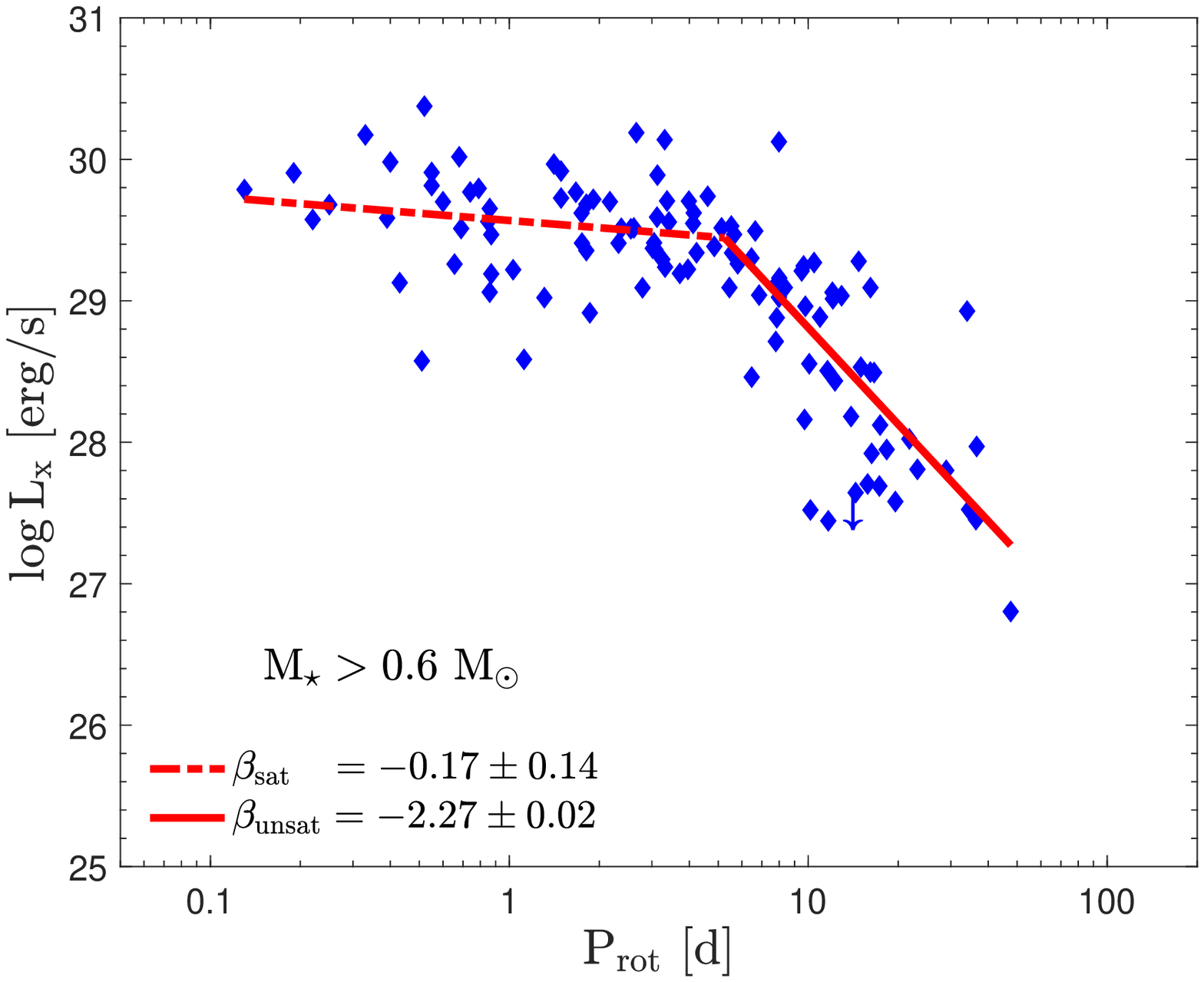}\par 
                \includegraphics[width=\linewidth]{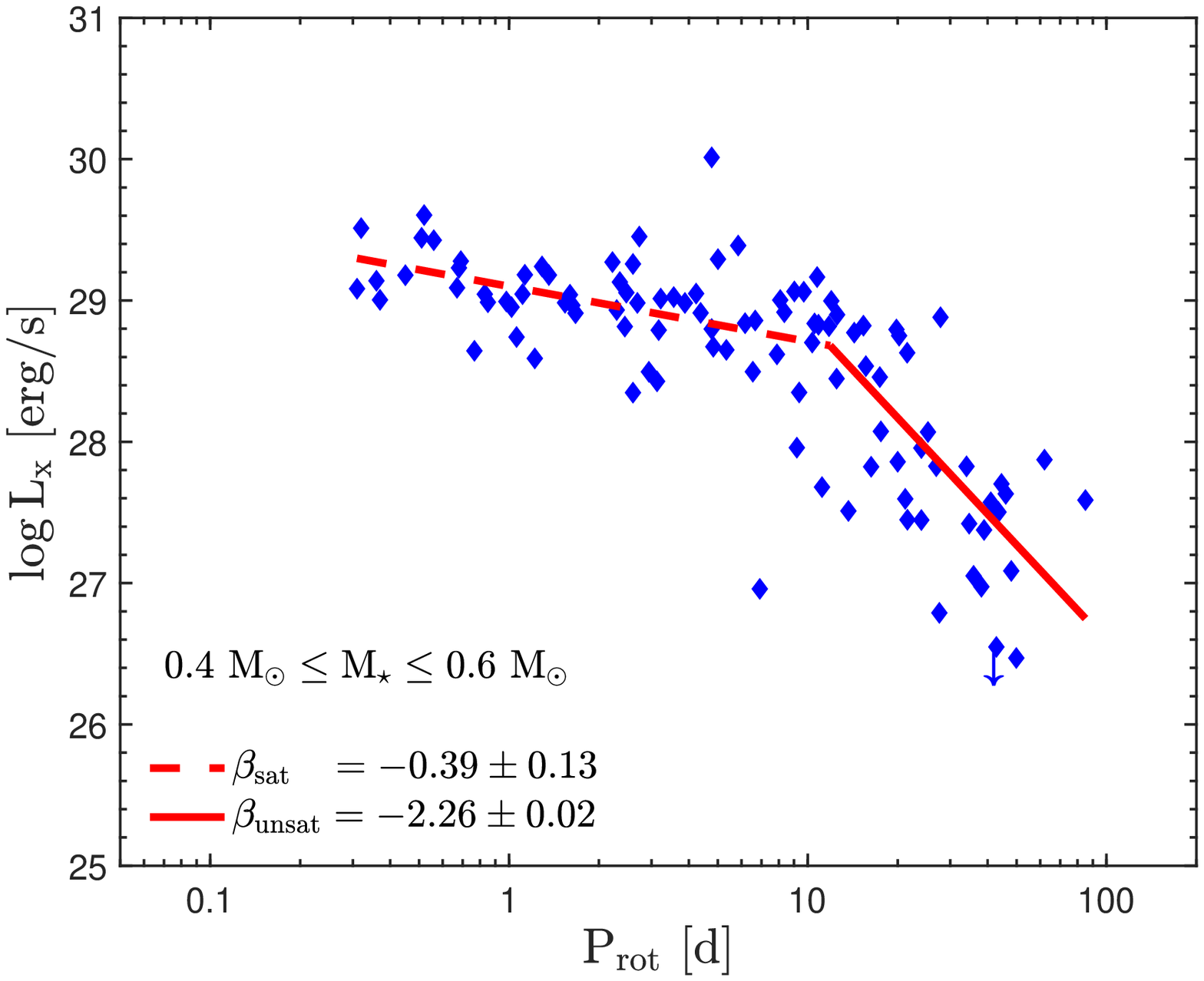}\par 
                \includegraphics[width=\linewidth]{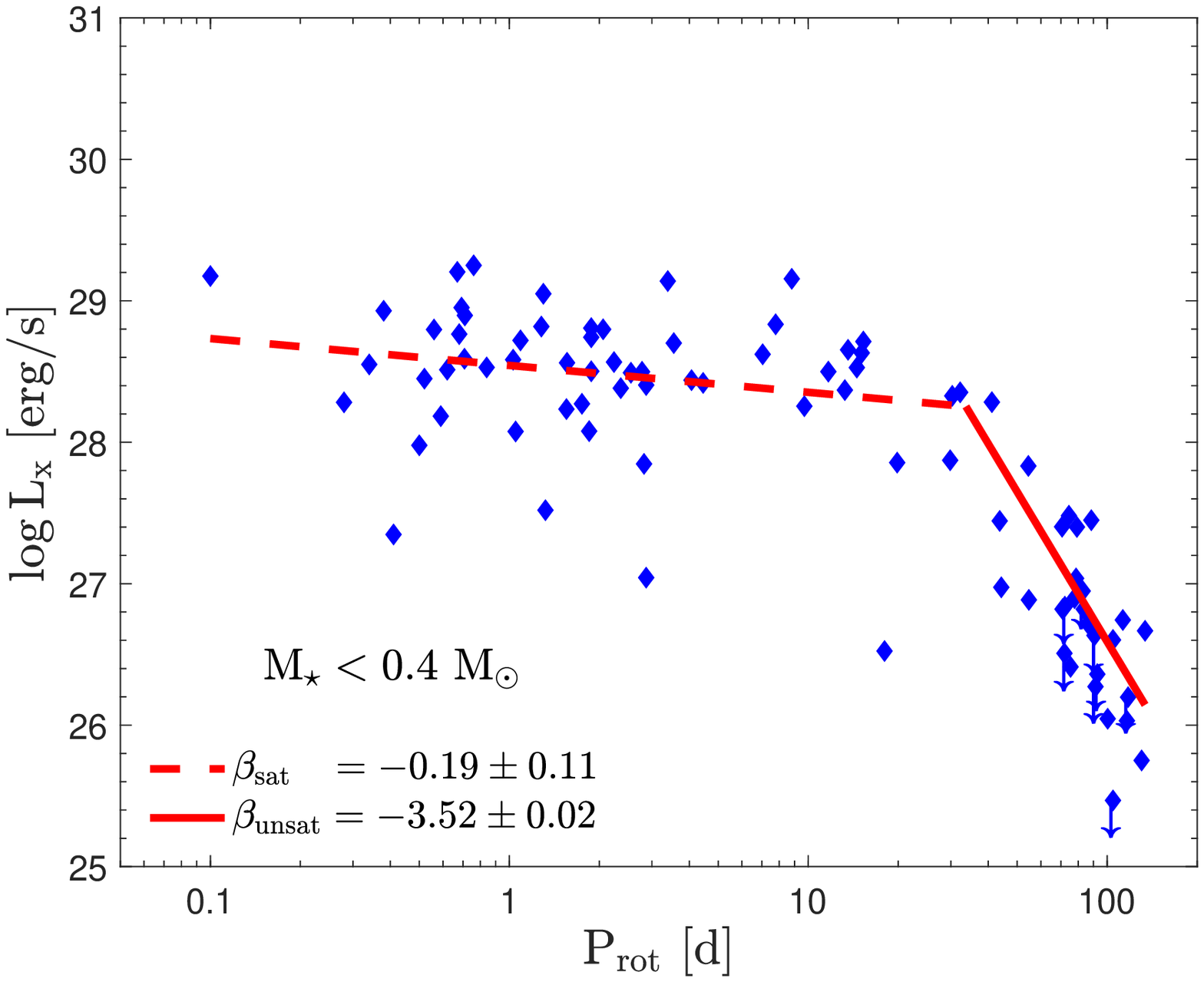}\par
                \includegraphics[width=\linewidth]{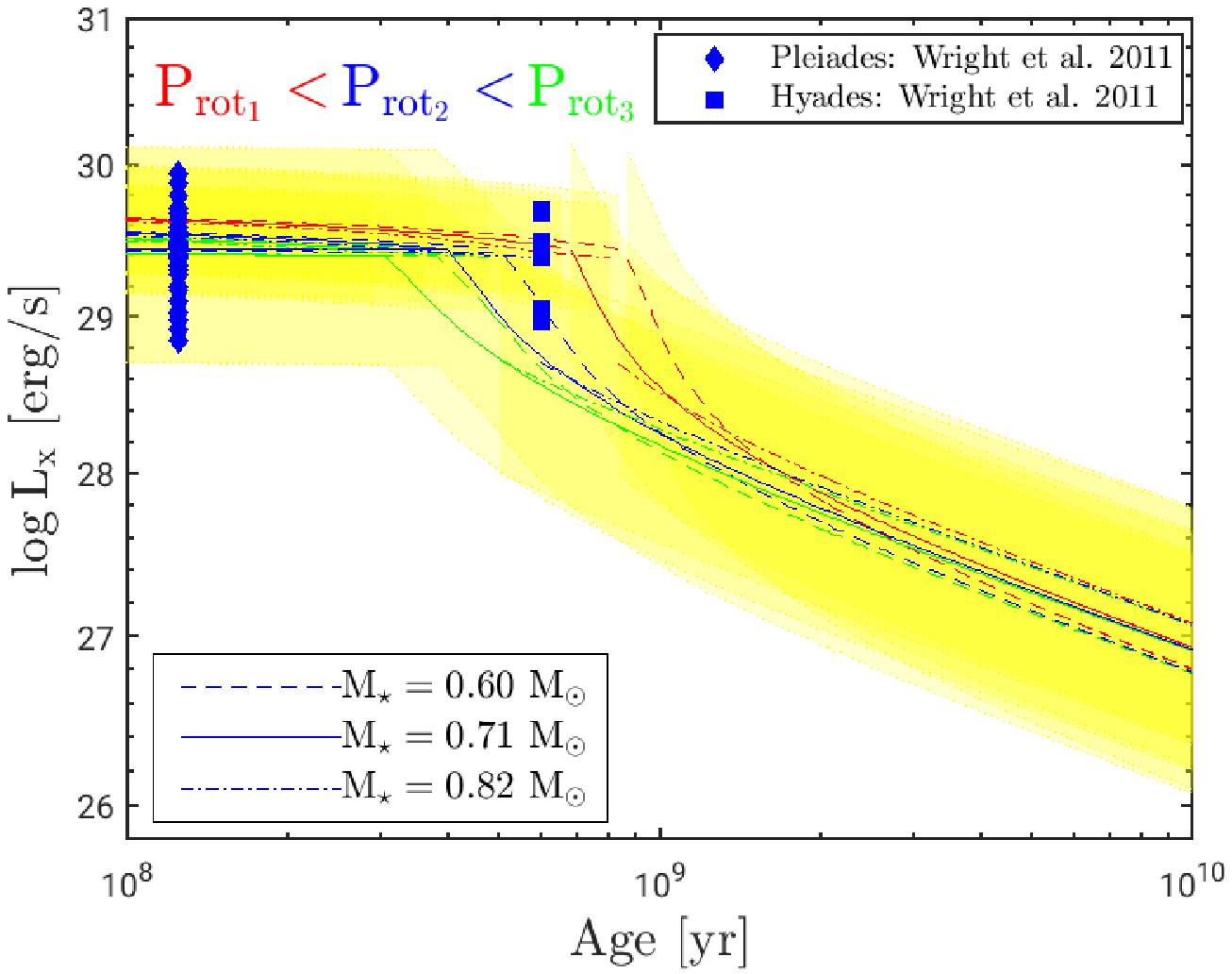}\par
                \includegraphics[width=\linewidth]{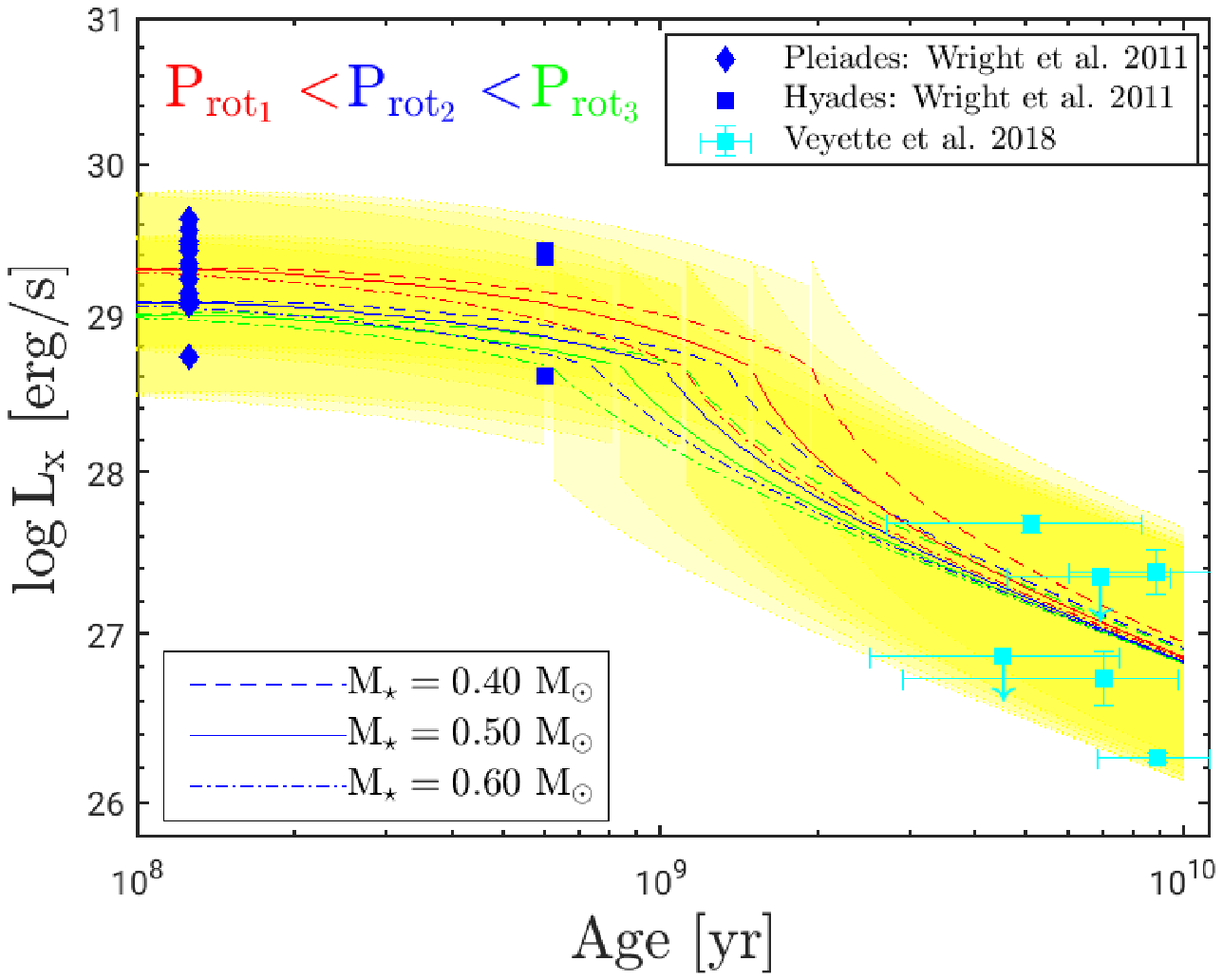}\par
                \includegraphics[width=\linewidth]{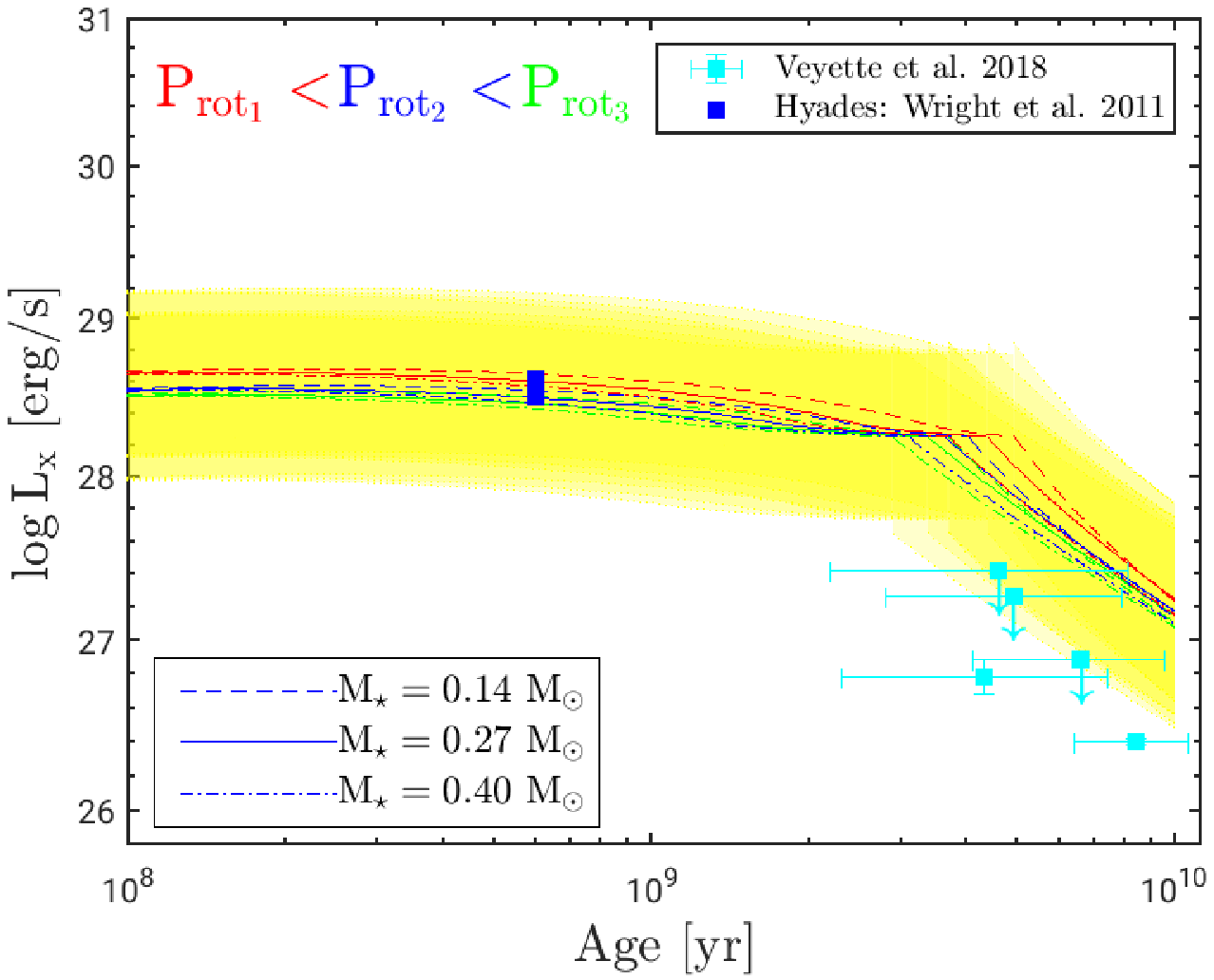}\par
        \end{multicols}
        \caption{\textbf{On the left}: Results of the activity-rotation relation fitting for the three mass ranges considered in this work. 
                \textbf{On the right}: Retrieved $L_{\rm x}$-age relation from angular evolution models for the same mass bins together with M~dwarfs with known ages from the literature by \citet{Wright2011} (blue points) and \citet{Veyette2018} (squared cyan). The yellow region shows the vertical $L_{\rm x}-$spread from the standard deviation of the observed $L_{\rm x}-P_{\rm rot}$ relation. Three different initial rotation periods are shown: $P_{\rm rot,1}=1.54$~d (red line), $P_{\rm rot,2}=5.51$~d (blue line), and $P_{\rm rot,3}=8.83$~d (green line).
                 In the predicted $L_{\rm x}-$age relation we show the model for the central mass bin (solid line), together with the lower and upper mass boundaries (dashed and dotted line, respectively).}
        \label{Act_rot_m}
\end{figure*}
\begin{table*}
        \caption{Results from fitting the relation of activity to rotation in $L_{\rm x}-P_{\rm rot}$ space for the full sample and three mass ranges (see Eq. \ref{fit_kevin} for more details).}  
        \begin{center}
                \begin{tabular}{cccccc}
                        \midrule[0.5mm]                    
                        \multicolumn{1}{c}{\textbf{Mass range}} 
                        &\multicolumn{1}{c}{\textbf{$N_{\star}$}} &\multicolumn{1}{c}{\textbf{$\beta_{\rm sat}$}}  
                        &\multicolumn{1}{c}{\textbf{$\beta_{\rm unsat}$}}
                        &\multicolumn{1}{c}{\textbf{$P_{\rm rot,sat}$}}
                        &\multicolumn{1}{c}{\textbf{$\log \left(L_{\rm x,sat}\right)$ ($P_{\rm rot}=1$ d)}}\\
                        
                        \multicolumn{1}{c}{}  
                        &\multicolumn{1}{c}{}  
                        &\multicolumn{1}{c}{} 
                        &\multicolumn{1}{c}{}
                        &\multicolumn{1}{c}{\textbf{[d]}}
                        &\multicolumn{1}{c}{\textbf{[erg/s]}}\\
                        
                        \midrule[0.5mm]
                        Full Sample&302&-0.14$\pm$0.10&-2.25$\pm$0.02&\phantom{0}8.5$\pm$1.0&29.11$\pm$0.11\\
                        $M_{\star} > 0.6M_{\odot}$&113&-0.17$\pm$0.14&-2.27$\pm$0.02&\phantom{0}5.2$\pm$0.7&29.56$\pm$0.13\\
                        $0.4M_{\odot}\leq M_{\star} \leq 0.6M_{\odot}$&102&-0.39$\pm$0.13&-2.26$\pm$0.02&11.8$\pm$2.0&29.10$\pm$0.16\\
                        $M_{\star} < 0.4M_{\odot}$&87&-0.19$\pm$0.11&-3.52$\pm$0.02&33.7$\pm$4.5&28.54$\pm$0.20\\
                        
                        \bottomrule[0.5mm]
                \end{tabular}
                \label{Fit_res}
        \end{center}
\end{table*}
Figure~\ref{Act_Prot} clearly shows systematic trends with stellar mass. In particular, the saturated $L_{\rm x}$ level decreases for lower $M_{\star}$ and the $P_{\rm rot,sat}$ turnover point is higher for lower $M_{\star}$. In order to search for differences in the activity-rotation relation of partially and fully convective stars, we therefore split the sample into three stellar mass ranges: lower, medium, and higher stellar masses.
We used the results from \citet{Jao_2018}, who assigned the transition to fully convective stars to $M_{\rm K_{s}} = 6.7$~mag (dashed black line in Fig.~\ref{Mks}); this corresponds to $V-J \approx 4$~mag. This approach is justified a posteriori by the fact that at $M_{\rm K_{s}} > 6.7$~mag there are mostly objects from \citet{Wright2016} and \citet{Wright2018}, where only M3 and later stars are included. Comparing Fig.~\ref{Mks} to the empirical relation between SpT and $V-J$ from \citet{Stelzer2016}, we found that $V-J = 4$~mag corresponds to SpT$\sim$M3.5.
Based on the comparison of $M_{\rm K_{s}}$, $V-J$, $M_{\star}$, and SpT, we therefore locate the fully convective transition at $M_{\star} = 0.4\hspace{1mm}M_{\odot}$. In order to split the full sample into three $M_{\star}$ bins, we considered our fully convective mass transition and then subdivided the partially convective sample into two mass ranges. In particular, the three stellar mass ranges are (1) $0.14\hspace{0.5mm}M_{\odot} \leq M_{\star}\leq 0.40\hspace{0.5mm}M_{\odot}$, (2) $0.40\hspace{0.5mm}M_{\odot} \leq M_{\star}\leq 0.60\hspace{0.5mm}M_{\odot}$, and (3) $0.60\hspace{0.5mm}M_{\odot} \leq M_{\star}\leq 0.82\hspace{0.5mm}M_{\odot}$. The number of stars in each mass bin is given in Table~\ref{Fit_res}. We recall that the highest mass bin also comprises late K-type stars.

We separately investigated the relation of activity to rotation in these three mass ranges by applying the same fitting procedure we used above. In Fig.~\ref{Act_rot_m} we show the results of our fitting analysis for the three mass ranges.
For each mass range the saturated regime has a nonconstant X-ray activity level. The slope $\beta_{\rm sat}$ for the high-mass range is flatter than the slope for the low-mass range. The slope in the intermediate-mass range is the steepest. Nonetheless, the slope in the saturated regime is independently detected at the 1$\sigma$ level in all three mass bins, which raises the statistical significance of this result above the 1$\sigma$ confidence in the global fit.
We confirmed the result found by \citet{Pizzolato2003} that the breaking point $P_{\rm rot,sat}$ moves to longer periods with decreasing stellar mass. 
In the unsaturated regime the slopes are similar ($\boldsymbol{\beta_{\rm unsat}\approx -2.2}$) for the higher and the intermediate mass range,  but the lowest mass range shows a much steeper decline of $L_{\rm x}$ with $P_{\rm rot}$ ($\boldsymbol{\beta_{\rm unsat,<0.4\,M_{\odot}}=-3.5}$).
In Table~\ref{Fit_res} the fit parameters are listed for all mass ranges, together with the X-ray luminosity $<L_{\rm x,sat}>$ calculated at $P_{\rm rot}=1$~d with the fit procedure. From this we see that the X-ray activity level in the saturated regime displays a continuous decrease toward later SpT \citep[also observed by][on a much smaller sample]{Stelzer2016}.

\subsection{X-ray luminosity vs age}
 X-ray activity and rotation are both known to undergo significant change during the stellar lifetime. $L_{\rm x}$ decays by a factor 1000 from the pre-main sequence (PMS) to the main-sequence (MS) \citep{Preibisch2005}, for instance, presumably because the dynamo efficiency decreases. The rotation periods are observed to decrease during PMS contraction, starting from an initially broad distribution with $P_{\rm rot} \sim 0.5 \text{ to }10$\,d \citep[depending on stellar mass;][]{Irwin2011}. The further evolution of the rotation rate during the 
 MS life is thought to be ruled by angular momentum loss mediated by magnetic winds \citep{Kawaler_1988}. Different wind models have been developed to predict the rotational evolution, see \citet{Matt_2015} and \citet{Garraffo2018} for M stars, for example. However, no theory exists that quantifies the decay in X-ray luminosity during the MS evolution, and the lack of field M dwarfs with known age has impeded an observational study of the $L_{\rm x}-$age relation for ages beyond that of open clusters such as the Hyades.
\subsubsection{Predicted relation of X-ray luminosity to age}
\label{Lx_constr}
Here we predict the time-evolution of the X-ray emission by combining the observed $L_{\rm x} - P_{\rm rot}$ relation from Sect.~\ref{LxProt_space} with the spin-down models  ($P_{\rm rot} -$ age) from \citet{Matt_2015}.
We perform this analysis individually for the three mass bins considered in Sect.~\ref{LxProt_space}. We calculated the rotation periods for stars with mass equal to the edges and the mean of the three mass bins, using the model of \citet{Matt_2015}, starting from an age of $5$\,Myr and evolving to an age of $10$\,Gyr. Because our observed $L_{\rm x} - P_{\rm rot}$ relations refer to a range  of masses (see Table~\ref{Fit_res}), we extracted the $P_{\rm rot}$ evolution from the angular momentum evolution model for the central mass of the bin, as well as for the mass of the lower and the upper boundaries. This allowed us to take the mass spread within each of our three mass bins into account. The rotational evolution depends on the initial rotation period ($P_{\rm rot,in}$) of the star, which is not a unique value (see our discussion above), as is known from observations in regions
of star formation. We therefore took this spread in the boundary conditions into account, and we investigated three different initial values for the rotation periods. These led to three tracks for given $M_{\star}$.

Our procedure for deriving an $L_{\rm x}-$age relation, carried out for each of the three mass bins from Sect.~\ref{LxProt_space} separately, is the following. 
We extracted for each mass value (the central mass of the bin, the lower and the upper mass boundaries) and ages from $5$\,Myr to $10$\,Gyr the rotation periods from the \citet{Matt_2015} model using $P_{\rm rot,in} = 1.54, 5.51$ and $8.83$\,d for the initial period. We thus obtained a total of nine tracks for the age evolution of the rotation period and three tracks for the three values of $P_{\rm rot,in}$, and this for each of three masses. We show these tracks in Fig.~\ref{ProtAge}. 
Then we calculated the $L_{\rm x}$ value corresponding to each $P_{\rm rot}$ value from the appropriate best-fit relation given in Table~\ref{Fit_res}.
To consider the vertical spread observed in the saturated and unsaturated regimes of $L_{\rm x}$ versus $P_{\rm rot}$ , we assign the observed $L_{\rm x}$ standard deviation to the constructed X-ray luminosities in each mass range. 

In the right panel of Fig.~\ref{Act_rot_m} we show our constructed $L_{\rm x}-$ age relation for each mass bin and for the three initial period values. 
The computed vertical spread is shown as the yellow region. The standard deviation in the predicted $L_{\rm x}$ is generally larger than the difference in tracks that is due to the different initial rates or different masses within the same mass bin.

\subsubsection{Comparison to observations}
\label{known_age}
\begin{table*}
        \caption{Relevant parameters for the sample of M dwarfs from \citet{Veyette2018}. $FLAG_{\rm D}$ are not listed because all {\em Gaia} parallaxes are reliable for this sample (see text in Sect.~\ref{known_age}).}            
        \begin{center}
                \begin{tabular}{lccccc}
                        \midrule[0.5mm]                    
                        \multicolumn{1}{c}{\textbf{Name}} 
                        &\multicolumn{1}{c}{\textbf{Age}}  
                        &\multicolumn{1}{c}{\textbf{D}}
                        &\multicolumn{1}{c}{\textbf{$M_{\star}$}}
                        &\multicolumn{1}{c}{\textbf{X-ray catalog}}
                        &\multicolumn{1}{c}{\textbf{$\log\left(L_{\rm x}\right)$}}\\

                        \multicolumn{1}{c}{}  
                        &\multicolumn{1}{c}{\textbf{[Gyr]}} 
                        &\multicolumn{1}{c}{\textbf{[pc]}}
                        &\multicolumn{1}{c}{\textbf{[$M_{\odot}$]}}
                        &\multicolumn{1}{c}{}
                        &\multicolumn{1}{c}{\textbf{[erg/s]}}\\

                        \midrule[0.5mm]
                        GJ 176&$8.8_{-2.8}^{+2.5}$&\phantom{0}9.473$\pm$0.006&0.47$\pm$0.02&RASS/FSC&27.38$\pm$0.13\\
                        GJ 179&$4.6_{-2.4}^{+3.5}$&12.360$\pm$0.009&0.33$\pm$0.01&RASS&<27.42\\
                        GJ 436&$8.9_{-2.1}^{+2.3}$&\phantom{0}9.755$\pm$0.008&0.43$\pm$0.01&3XMM-DR8&26.26$\pm$0.02\\
                        GJ 536&$6.9_{-2.3}^{+2.5}$&14.412$\pm$0.009&0.56$\pm$0.01&RASS&<27.35\\
                        GJ 581&$6.6_{-2.5}^{+2.9}$&\phantom{0}6.299$\pm$0.002&0.40$\pm$0.01&RASS&<26.88\\
                        GJ 617A&$5.1_{-2.4}^{+3.2}$&10.767$\pm$0.003&0.58$\pm$0.02&RASS/BSC&27.68$\pm$0.05\\
                        GJ 625&$7.0_{-4.1}^{+2.7}$&\phantom{0}6.473$\pm$0.001&0.50$\pm$0.02&RASS/FSC&26.74$\pm$0.16\\
                        GJ 628&$4.3_{-2.0}^{+3.1}$&\phantom{0}4.306$\pm$0.001&0.33$\pm$0.01&RASS/FSC&26.78$\pm$0.10\\
                        GJ 649&$4.5_{-2.0}^{+3.0}$&10.382$\pm$0.003&0.51$\pm$0.21&RASS/FSC&<26.87\\
                        GJ 849&$4.9_{-2.1}^{+3.0}$&\phantom{0}8.802$\pm$0.003&0.40$\pm$0.01&RASS/FSC&<27.27\\
                        GJ 876&$8.4_{-2.0}^{+2.2}$&\phantom{0}4.675$\pm$0.001&0.31$\pm$0.01&3XMM-DR8&26.40$\pm$0.01\\

                        \bottomrule[0.5mm]
                \end{tabular}
                \label{Veye_par}
        \end{center}
\end{table*}
The validity of the relation for the age-decay of X-ray luminosity that we constructed from the spin-down models and the empirical relation of X-ray activity to rotation can be tested with stars of known age and X-ray luminosity. Such samples are notoriously sparse in the M-dwarf regime. We took three such samples from the literature: M stars in the Pleiades ($125$\,Myr), in the Hyades ($\sim 600$\,Myr), and field M dwarfs with individual ages determined from a chemo-kinematic study. The X-ray luminosities of these objects are plotted over the predicted $L_{\rm x}-$age relation in the right panel in Fig.~\ref{Act_rot_m}.

The X-ray luminosities of the Pleiades and Hyades werere extracted from \citet{Wright2011}, observed in the $0.1-2.4$\,keV band. \citet{Wright2011} provide a list of stars with detected X-ray emission for both clusters, but no upper limits for stars with X-ray luminosities below the detection limit. They computed $L_{\rm x}$ with the adopted distance equal to $133$\,pc for stars in the Pleiades and $46$\,pc for stars in the Hyades.

Field stars were taken from \citet{Veyette2018}, who determined individual ages for $11$ nearby planet-hosting M dwarfs using a combination of kinematics and chemical evolution. First we calculated the distances and stellar parameters for these stars as described in Sect.~\ref{sample} for the rotation-activity sample. Then we searched for X-ray detections of these stars in the {\em ROSAT} FSC \citep{Voges2000}, the {\em ROSAT} BSC \citep{Voges1999}, and the {\em 3XMM-DR8} catalog \citep{Rosen2016}. For the $\text{six}$ stars that we were able to identify with a source in one of the above catalogs, we derived the X-ray luminosities from the cataloged count rates in the same manner in which we treated the stars from \citet{Gonzalez-Alvarez2019} (see Sect.~\ref{stellar_par}). For the remaining $\text{five}$ stars, we estimated the upper limit on $L_{\rm x}$ based on the {\em ROSAT} All-Sky Survey (RASS). Specifically, we extracted the RASS exposure time at the location of each of the stars from the exposure maps available from the {\em ROSAT} webpage at the Max-Planck-Institut f\"ur extraterrestrische Physik\footnote{\url{http:xray.mpe.mpg.de/cgi-bin/rosat/rosat-survey}}. Then we obtained the individual upper limit count rates from the estimated RASS sensitivity limit shown as lower envelope in the plot that shows the RASS count rate versus exposure time in Fig.~4 of \citet{Stelzer2013}. In Table~\ref{Veye_par} we list all derived parameters for the stars from \citet{Veyette2018} that are relevant for our purpose.
These are the age with its uncertainty extracted from \citet{Veyette2018} (Col. 2), the adopted distance derived from the quality criteria as defined in Sect.~\ref{stellar_par} (Col. 3, all {\em Gaia} distances are reliable for this sample, $FLAG_{\rm D}= 1 1$), the stellar mass (Col. 4), and the X-ray instrument and luminosity in the $0.1-2.4$\,keV band (Cols. 5 and 6). 

Each star from \citet{Veyette2018} is plotted on the right side of Fig.~\ref{Act_rot_m} in the respective panel corresponding to its individual stellar mass. 
Pleiades and Hyades stars from \citet{Wright2011} are plotted in Fig.~\ref{Act_rot_m}, using the stellar masses they computed using $V-K_{\rm s}$ magnitudes.

Our constructed $L_{\rm x}$-age relation places the Pleiades M dwarfs in the saturated regime and the Hyades M dwarfs as well, except for the highest mass bin ($M_{\star} = 0.6\text{ to }0.8\,{\rm M_\odot}$), where the age of the Hyades corresponds to the turnover point between saturated and unsaturated regime. Interestingly, both open clusters span the full spread of X-ray luminosities inferred from the \citet{Matt_2015} rotational evolution tracks and the observed $L_{\rm x} - P_{\rm rot}$ relation of field M dwarfs (yellow region in Fig.~\ref{Act_rot_m}). 

The field M dwarfs from \citet{Veyette2018} span ages from $\sim 4\text{ to }9$\,Gyr and masses within the range of our two lower-mass bins. At a given age and mass, this sample presents a spread in X-ray luminosity of more than an order of magnitude. None of these stars has $M_{\star}>0.6\,{\rm M_{\odot}}$, corresponding to our high-mass bin. While the intermediate-mass stars ($M_{\star} = 0.4-0.6\,{\rm M_\odot}$) fall within the predicted range of $L_{\rm x}$ (yellow zone in Fig.~\ref{Act_rot_m}), the fully convective stars ($M_{\star} < 0.4\,{\rm M_{\odot}}$) are clearly underluminous with respect to the prediction of our $L_{\rm x}$-age relation. 
Our procedure  overpredicts the X-rays of the fully convective field M dwarfs because the spin-down model provides rotation periods that are faster than observed for these stars \citep[as noted in ][]{Matt_2015}.
\subsection{Fractional X-ray luminosity versus Rossby number}
 \label{LxLbolRo_space}
 As described in Sect.~\ref{stellar_par}, we computed the $\tau_{\rm conv}$ for the full sample using the relation by \citet{Wright2018}, which is valid over the range $1.1<V-K_{\rm s}<7.0$. To investigate if this empirical $\tau_{\rm conv}$ scale is consistent with theoretical values, we also constructed the $L_{\rm x}/L_{\rm bol}$-$R_{\rm O}$ relation for the $\tau_{\rm conv}$ parameterizations of \citet{Cranmer_2011} and \citet{Brun2017}. The relation between $\tau_{\rm conv}$ and $T_{\rm eff}$ of \citet{Cranmer_2011} is the result of a parameterized fit of 1D stellar structure models. \citet{Brun2017} presented the fluid Rossby number as a function of $M_{\star}$ and $\Omega_{\star}$ computed with 3D stellar models based on mixing length theory, and \citet{Wright2018} derived $\tau_{\rm conv}$ empirically as a function of $V-K_{\rm s}$ color from a study of the $L_{\rm x}/L_{\rm bol}-R_{\rm O}$ relation for fully convective stars. Because the relations from \citet{Cranmer_2011} and \citet{Brun2017} are calibrated only for partially convective stars, we excluded from the analysis in $L_{\rm x}/L_{\rm bol}$ - $R_{\rm O}$ space all stars with $T_{\rm eff}<3300$~K, corresponding to $M_{\star}<0.4\rm {M_{\odot}}$. 
 \begin{table}[t!]
        \begin{minipage}[r]{\linewidth}
                \centering
                \includegraphics[width=\textwidth]{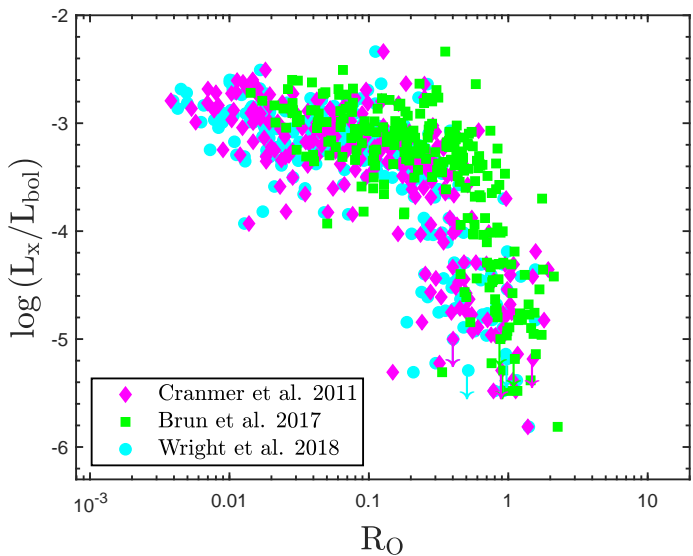}
                \captionof{figure}{Comparison of $L_{\rm x}/L_{\rm bol}$ vs. $R_{\rm O}$ obtained for the three $\tau_{\rm conv}$ parameterizations by \citet{Cranmer_2011} (purple filled rhombus), \citet{Brun2017} (green filled square), and \citet{Wright2018} (cyan filled circle).}
                \label{tau_PLOTcomp}
        \end{minipage}
        \caption{Results from the fitting procedure applied to the $L_{\rm x}/L_{\rm bol}$ vs. $R_{\rm O}$ relation shown in Fig.~\ref{tau_PLOTcomp}, computed for the three $\tau_{\rm conv}$ parameterizations described in Sect.~\ref{LxLbolRo_space}.}
        \begin{minipage}{0.5\linewidth}
                \begin{center}
                        \begin{tabular}[l]{lccc}
                                \midrule[0.5mm]  
                                \multicolumn{1}{c}{Parameterization}                                       
                                &\multicolumn{1}{c}{\textbf{$\beta_{\rm sat}$}} 
                                &\multicolumn{1}{c}{\textbf{$\beta_{\rm unsat}$}}
                                &\multicolumn{1}{c}{\textbf{$R_{\rm O,\rm sat}$}}\\
                                
                                \multicolumn{1}{c}{}  
                                &\multicolumn{1}{c}{}
                                &\multicolumn{1}{c}{} 
                                &\multicolumn{1}{c}{}\\
                                
                                \midrule[0.5mm]
                                Cranmer et al. 2011&-0.31$\pm$0.08&-2.03$\pm$0.01&0.19$\pm$0.02\\
                                Brun et al. 2017&-0.33$\pm$0.08&-2.92$\pm$0.02&0.41$\pm$0.02\\
                                Wright et al. 2018&-0.21$\pm$0.08&-1.99$\pm$0.01&0.14$\pm$0.01\\
                                
                                \bottomrule[0.5mm]
                        \end{tabular}
                \end{center}
        \end{minipage}\hfill
        \label{res_tau_comp}
 \end{table}
 It is important to note that the different scaling laws for $\tau_{\rm conv}$ result in different values for the Sun. Therefore we normalized the three relations in order to obtain a fixed solar value. 
 We scaled the \citet{Cranmer_2011} and \citet{Brun2017} parameterization to the one by \citet{Wright2018}, taking the ratio between $\tau_{\rm conv}$ for the Sun as normalization factor.
 In particular, we computed the solar $\tau_{\rm conv}$ by \citet{Cranmer_2011} using $T_{\rm eff,\odot}= 5778$~K \citep{Brandenburg2017} and the solar $\tau_{\rm conv}$ by \citet{Wright2018} with $(V-K)_{\odot}=1.5$ mag\footnote{\url{http://mips.as.arizona.edu/~cnaw/sun.html}}. 
 Because \citet{Brun2017} provide $R_{\rm O}$ already normalized to the solar mass and rotation rate, we computed the solar $\tau_{\rm conv}$ using $P_{\rm rot,\odot} = 24.5$ d \citep{Brandenburg2017}.
 In Table~\ref{XRay_Results} and \ref{ALLdata_par} we list for each star the three $R_{\rm O}$ values from the different relations for $\tau_{\rm conv}$ . In Fig.~\ref{tau_PLOTcomp} we show how $L_{\rm x}/L_{\rm bol}$ versus $R_{\rm O}$ changes according to the different adopted $\tau_{\rm conv}$ parameterizations. The $L_{\rm x}/L_{\rm bol}$ - $R_{\rm O}$ relation with the \citet{Wright2018} and \citet{Cranmer_2011} $\tau_{\rm conv}$ values are similar, but the \citet{Brun2017} parameterization is shifted toward higher $R_{\rm O}$ values and has a smaller spread in the unsaturated regime.

We applied the fitting procedure used in Sect.~\ref{LxProt_space} to the three $L_{\rm x}/L_{\rm bol}$ versus $R_{\rm O}$ relations. The best-fit parameters are shown in Table~\ref{res_tau_comp}. All three relations yield a nonconstant saturated level, that is, a decrease in $L_{\rm x}/L_{\rm bol}$ for higher Rossby numbers, at the $3~\sigma$ level.
The $R_{\rm O,sat}$ for the \citet{Brun2017} parameterization has a noticeably larger breaking point than those from \citet{Cranmer_2011} and \citet{Wright2018} calibrations. 
Moreover, the slope in the unsaturated regime from \citet{Brun2017} is steeper than the other two, showing a much more abrupt activity decrease toward higher $R_{\rm O}$.
Another interesting result is the visible and remarkable double gap around $0.2\leq R_{\rm O}\leq1.2$ , which corresponds to $-4.5\leq~L_{\rm x}/L_{\rm bol}\leq -3.5$ (see Fig.~\ref{tau_PLOTcomp})

\section{Discussion}
\label{disc}
We presented a thorough investigation of the shape of the relation of X-ray activity$ \text{to }$rotation in $L_{\rm x}-P_{\rm rot}$ space for a comprehensive sample of M dwarfs, and we studied for the first time, to our knowledge,  the effect of the $\tau_{\rm conv}  $ calibration on this relation in $L_{\rm x}/L_{\rm bol}-R_{\rm O}$ space.
We created the largest and most homogeneous database of rotational periods and X-ray activity for field M dwarfs to date by taking new observations with {\em XMM-Newton} and {\em Chandra} satellites and updating data from the literature.
We computed stellar parameters from calibrated photometry for a total of 302 stars with measured $P_{\rm rot}$ and $L_{\rm x}$, including {\em Gaia} parallaxes when reliable according to \citet{Lindegren2018}, and our own criteria.
With our combination and homogenization analysis of the whole sample originating from previous studies, we reduced possible observational biases caused by the different limitations of the samples from the literature. In particular, because \citet{Wright2011}, \citet{Stelzer2016} and \citet{Gonzalez-Alvarez2019} used X-ray data from the archives without including upper limits in the analysis, this leads to a bias toward X-ray bright stars. On the other hand, \citet{Wright2016} and \citet{Wright2018} selected stars from the MEarth project, where only fully convective stars with long $P_{\rm rot}$ are included. Our new sample with deep dedicated X-ray observations for 14 {\em K2}-selected M dwarfs avoids the X-ray brightness bias, but is limited to 14 stars.

We analyzed the $L_{\rm x}-P_{\rm rot}$ relation by applying a two-slope power-law fit in three different mass ranges. Next to the known two-regime behavior with saturation for fast-rotating stars and decreasing $L_{\rm x}$ for higher $P_{\rm rot}$ above a certain threshold, we find that the $L_{\rm x}$ level in the saturated regime is not constant, but decreases slightly with increasing $P_{\rm rot}$. In the saturated regime the lowest mass stars have the lowest X-ray luminosities, showing a large ($\approx$2~dex) $\rm L_{x}$ spread (see the bottom panel of Fig.~\ref{Act_Prot} ). Here we likely probe the rotation-independent decrease in X-ray emission in late-M dwarfs that has been ascribed to poor coupling between matter and magnetic field in the increasingly neutral atmospheres at the bottom of the MS and the ensuing shut-off of activity \citep{Mohanty2002} probably caused by the increasing electrical resistivity in such cool atmospheres. 
We confirmed past results by \citet{Pizzolato2003} for which the breaking point between the two power laws occurs at higher $P_{\rm rot}$ as $M_{\star}$ decreases. However, in our several times larger sample, the values we find for the turnover points are much higher than those presented in their historical study. 

The nonconstant X-ray emission level in the saturated regime was first noted by \citet{Reiners2014} in terms of $L_{\rm x}/L_{\rm bol}$. As possible explanation, they suggested a property of the dynamo or a residual mass-dependence in the saturated regime. We see the negative slope in each of the three mass bins we examined (see, e.g., the left panel in Fig.~\ref{Act_rot_m}). The likely cause therefore is some rotation dependence of the dynamo even for these fast rotators.
We measured a steeper slope in the unsaturated regime for stars with $M_{\star}<0.4\,\rm{M_{\odot}}$ (fully convective stars). 

We used our best-fit parameters of the activity-rotation relation to construct the $L_{\rm x}-$age relation using spin-down models by \citet{Matt_2015}. In the time-evolution tracks for a given narrow mass range at a certain mass-dependent point, the evolution of different initial periods starts to diverge. This is not visible in the $L_{\rm x}-$age relation as long as the stars remain saturated, but when they drop out of saturation and $L_{\rm x}$ starts to decrease, the tracks with different initial periods also diverge in $L_{\rm x}-$age space. However, the range of our predicted $L_{\rm x}$ for different initial $P_{\rm rot}$ at given age and mass (i.e., the tracks in the right panel in Fig.~\ref{Act_rot_m}) is much smaller than the $L_{\rm x}-$spread we inferred from the observed relation of X-ray activity to rotation (yellow region in Fig.~\ref{Act_rot_m}). Therefore we cannot distinguish the X-ray evolution of stars with different initial rotation periods.

By comparing our constructed $L_{\rm x}-$age relation to stars with known age, we found that the Hyades stars in our high-mass bin ( $0.6-0.8\,{\rm M_\odot}$; corresponding to late-K to early-M SpT) are located at the onset of the unsaturated regime in the $L_{\rm x}-$age relation. The rotation periods of the Hyades in that mass range are \citep[$\sim$10$-$20\,d;][]{Douglas_2019}, which is roughly consistent with the $P_{\rm rot}$ at which the transition from the saturated to the unsaturated regime takes place. Moreover, our result shows that the Pleiades and Hyades stars span the full range of $L_{\rm x}$ in the saturated regime, suggesting that the scatter of X-ray luminosity at a given rotation period has no evolutionary component from the zero-age MS onwards.
Because the model by \citet{Matt_2015} fails to produce the known long rotation periods ($\gnsim$50\,d) for M dwarfs, our model overpredicts the X-ray luminosity of fully convective field stars in our constructed relation of X-ray$  \text{ to }$age. This explains why in the fully convective bin the field M stars with gigayear-ages are located below the $L_{\rm x}$ expected from our relation.

We investigated for the first time, to our knowledge, how adopting different $\tau_{\rm conv}$ parameterizations can affect the shape of the $L_{\rm x}/L_{\rm bol}-R_{\rm O}$ relation. We performed this comparison for the Rossby numbers of \citet{Cranmer_2011}, \citet{Brun2017}, and \citet{Wright2018}, which originate from different approaches and have different ranges of validity.
In particular, the relation of \citet{Cranmer_2011} is valid only for stars with $T_{\rm eff}\geq 3300$~K, \citet{Brun2017} can be applied to a wide range of stellar masses (from $0.4~\rm M_{\odot}$ to $1.2~\rm M_{\odot}$), and the relation of \citet{Wright2018} is valid over 1.1~mag~$<V-K_{\rm s}<$~7.0~mag. Therefore, the stars we considered in our investigation of $L_{\rm x}/L_{\rm bol}-R_{\rm O}$ are the 242 out of the full sample that fulfill all these conditions. 

We applied the same fitting procedure used for the $L_{\rm x}-P_{\rm rot}$ relation to $L_{\rm x}/L_{\rm bol}-R_{\rm O}$ , and we identified the following interesting results: (1) all calibrations provide a decrease in $L_{\rm x}/L_{\rm bol}$ in the saturated regime, (2) the parameterization by \citet{Brun2017} yields a much steeper $\beta_{\rm unsat}$ slope with a breaking point at higher $R_{\rm O}$ than those from \citet{Cranmer_2011} and \citet{Wright2018} parameterizations, and (3) there is a remarkable double gap in $L_{\rm x}/L_{\rm bol}-R_{\rm O}$ space with a scarcity of objects slightly above and below $L_{\rm x}/L_{\rm bol}\sim10^{-4}$ (e.g., the right panel in Fig.~\ref{LxLbolRo_Ref}) that is difficult to explain as an observational bias.

As discussed above, \citet{Reiners2014} previously observed a slope in the saturated regime in $L_{\rm x}/L_{\rm bol}-R_{\rm O}$. They examined a sample in a broad mass range ($M_{\star} \leq1.4\rm M_{\odot}$) and found a slope $\beta_{\rm sat} = -0.16$. In our M dwarf sample we find a slightly steeper $\beta_{\rm sat}$ slope for all three $\tau_{\rm conv}$ parameterizations (see Table~\ref{res_tau_comp}). While above, based on $L_{\rm x}-P_{\rm rot}$, we argued that the existence of this slope is not a mass effect (because we see it in different mass bins), its actual steepness may depend on mass.

For the slope in the unsaturated regime, $\beta_{\rm unsat}$, we find significantly different results from the three Rossby parameterizations. \citet{Wright2011} studied the $L_{\rm x}/L_{\rm bol}-R_{\rm O}$ relation for partially convective stars, finding $R_{\rm O,sat}=0.16\pm0.03$, $\beta_{\rm unsat}= -2.7$, and $\left(L_{\rm x}/L_{\rm bol}\right)_{\rm sat}=-3.13$. When the \citet{Cranmer_2011} and \citet{Wright2018} calibrations are used, our slope in the unsaturated regime is substantially smaller than that value ($\beta_{\rm unsat}\simeq-2$), while \citet{Brun2017} yields a significantly larger slope ($\beta_{\rm unsat}=-2.9$). This latter parameterization also yields a much higher value for the break-point $R_{\rm O,sat}$ than \citet{Cranmer_2011} and \citet{Wright2018} and than the historical result by \citet{Wright2011}. The $L_{\rm x}/L_{\rm bol}-$Rossby relation constructed with the \citet{Brun2017} Rossby numbers visibly produces the smallest scatter of the data points, suggesting that it best represents the presumed universal mass-dependent parameter that rules the activity-rotation relation, and which is usually identified with the convective turnover time.

We speculate that the remarkable gap we found in $L_{\rm x}/L_{\rm bol}-R_{\rm O}$ space might be associated with a phase of stalled rotational evolution followed by an episode of rapid spin-down, which has recently been discussed in rotation studies of open clusters and solar-type field stars by \citet{Curtis2019} and \citet{Metcalfe2019}. In these works 
the rotation-age relation is studied for G- and K-type stars. According to these studies, stalling seems to last longer for lower stellar masses. This period stalling might lead to a pile-up of objects before the transition to the unsaturated regime, and combined with subsequent rapid spin-down, a gap around the breaking point in the relation of activity to rotation. We clearly see two such gaps in the $L_{\rm x}/L_{\rm bol}-R_{\rm O}$ space. Moreover, the bottom panel of Fig.~\ref{Act_Prot} demonstrates that this gap is present in different masses. If the evolution of the rotation period is responsible for these gaps, we would expect to see them in $L_{\rm x}-P_{\rm rot}$ space. There is some evidence for two sparsely populated regions around $\log L_{\rm x }\sim 28.2$ erg/s and $\log L_{\rm x}\sim 27.2$ erg/s. The upper region occurs at periods of $P_{\rm rot} \sim 10..30$~d, corresponding to the period gap in the $M_{\star}-P_{\rm rot}$ diagram of large samples from the {\em Kepler} mission \citep{Mcquillan2014}. While this coincidence is intriguing, the ``X-ray gap'' and the search for its origin require further investigation.

The relation of activity to rotation can be also studied by analyzing the emission of typical chromospheric spectral lines. For instance, \citet{Newton2017} analyzed the activity from the $\rm H_{\alpha}$ emission in $\rm L_{H_{\alpha}}/L_{bol}-R_{O}$ space by applying a broken power-law fit. They calculated the Rossby numbers with the $\rm \tau_{conv} $ parameterization from \citet{Wright2011}, therefore their results are not directly comparable with ours.

\section{Conclusions and outlook}
\label{Concl}
The collected and updated database of this work reduced the observational biases in the relation of X-ray activity to rotation. This leads to a series of interesting results, including (1) a nonconstant saturated level of the X-ray emission, (2) a significant steepening of the slope in the unsaturated regime for fully convective stars, (3) possible ``regions of avoidance'' in the $L_{\rm x}$ and $L_{\rm x}/L_{\rm bol}$ distribution that might be related to a discontinuous period evolution, and (4) the dependence of the shape of the $L_{\rm x}/L_{\rm bol}-R_{\rm O}$ relation on the assumptions made for the convective turnover time. We moreover predicted for the first time the evolution of M-dwarf X-ray emission for ages beyond $\sim 600$\,Myr, that is, after the stars drop out of saturation.  A focus of future studies should be the transition between saturated and unsaturated regimes of the rotation-activity relation, which is crucial for anchoring the dual power-law fit and to quantify the ``X-ray gap''. Unprecedentedly large samples can be expected from the All-Sky missions {\em TESS} and {\em eROSITA}, which yield $P_{\rm rot}$ up to 20~d and X-ray measurements 20 times deeper than {\em ROSAT}. These missions will be particularly useful to address these questions.

\begin{acknowledgements} 
        EM was supported by the \textsl{Bundesministerium f\"{u}r Wirtschaft und Energie} through the \textsl{Deutsches Zentrum f\"{u}r Luft- und Raumfahrt e.V. (DLR)} under grant number FKZ 50 OR 1808.
        SPM is supported by the European Research Council, under the European Union’s Horizon 2020 research and innovation program (agreement No. 682393, AWESoMeStars).
        AS’s work is supported by the STFC grant no. ST/R000824/1.
        This research made use of observations obtained with XMM-Newton, an ESA science mission with instruments and contributions directly funded by ESA Member States and NASA.
        The scientific results reported in this article are also based on observations made by the Chandra X-ray Observatory.
        We would especially like to thank A. Vanderburg for his public release of the analysed {\em K2} light curves. Funding for the {\em K2} mission is provided by the NASA Science Mission directorate.
        We also thank the anonymous referee for useful suggestions.
\end{acknowledgements}
\newpage
\bibliographystyle{aa} 
\bibliography{bibliography_AA201937408}

\begin{thebibliography}{50}
\expandafter\ifx\csname natexlab\endcsname\relax\def\natexlab#1{#1}\fi

\bibitem[{{Ag{\"u}eros} {et~al.}(2018){Ag{\"u}eros}, {Bowsher}, {Bochanski},
  {Cargile}, {Covey}, {Douglas}, {Kraus}, {Kundert}, {Law}, {Ahmadi}, \&
  {Arce}}]{Agueros2018}
{Ag{\"u}eros}, M.~A., {Bowsher}, E.~C., {Bochanski}, J.~J., {et~al.} 2018,
  \apj, 862, 33

\bibitem[{{Akritas} {et~al.}(1995){Akritas}, {Murphy}, \&
  {Lavalley}}]{Akritas1995}
{Akritas}, M.~G., {Murphy}, S.~A., \& {Lavalley}, M.~P. 1995, JASA, 90

\bibitem[{Arenou {et~al.}(2018)Arenou, Luri, Babusiaux, Fabricius, Helmi,
  Muraveva, Robin, Spoto, Vallenari, Antoja, Cantat-Gaudin, Jordi, Leclerc,
  Reyl{\'{e}}, Romero-G{\'{o}}mez, Shih, Soria, Barache, Bossini, Bragaglia,
  Breddels, Fabrizio, Lambert, Marrese, Massari, Moitinho, Robichon, Ruiz-Dern,
  Sordo, Veljanoski, Eyer, Jasniewicz, Pancino, Soubiran, Spagna, Tanga, Turon,
  \& Zurbach}]{Arenou2018}
Arenou, F., Luri, X., Babusiaux, C., {et~al.} 2018, A{\&}A, 616, A17

\bibitem[{{Boller, Th.} {et~al.}(2016){Boller, Th.}, {Freyberg, M. J.},
  {Tr\"umper, J.}, {Haberl, F.}, {Voges, W.}, \& {Nandra, K.}}]{Boller2016}
{Boller, Th.}, {Freyberg, M. J.}, {Tr\"umper, J.}, {et~al.} 2016, A\&A, 588,
  A103

\bibitem[{Bopp \& Evans(1973)}]{Bopp1973}
Bopp, B.~W. \& Evans, D.~S. 1973, MNRAS, 164, 343

\bibitem[{Brandenburg {et~al.}(2017)Brandenburg, Mathur, \&
  Metcalfe}]{Brandenburg2017}
Brandenburg, A., Mathur, S., \& Metcalfe, T.~S. 2017, \apj, 845, 79

\bibitem[{Brun {et~al.}(2017)Brun, Strugarek, Varela, Matt, Augustson, Emeriau,
  DoCao, Brown, \& Toomre}]{Brun2017}
Brun, A.~S., Strugarek, A., Varela, J., {et~al.} 2017, \apj, 836, 192

\bibitem[{{Chabrier} \& {K\"uker}(2006)}]{Chabrier2006}
{Chabrier} \& {K\"uker}. 2006, A\&A, 446, 1027

\bibitem[{Cranmer \& Saar(2011)}]{Cranmer_2011}
Cranmer, S.~R. \& Saar, S.~H. 2011, ApJ, 741, 54

\bibitem[{{Curtis} {et~al.}(2019){Curtis}, {Ag{\"u}eros}, {Douglas}, \&
  {Meibom}}]{Curtis2019}
{Curtis}, J.~L., {Ag{\"u}eros}, M.~A., {Douglas}, S.~T., \& {Meibom}, S. 2019,
  \apj, 879, 49

\bibitem[{{Douglas} {et~al.}(2014){Douglas}, {Ag{\"u}eros}, {Covey}, {Bowsher},
  {Bochanski}, {Cargile}, {Kraus}, {Law}, {Lemonias}, {Arce}, {Fierroz}, \&
  {Kundert}}]{Douglas2014}
{Douglas}, S.~T., {Ag{\"u}eros}, M.~A., {Covey}, K.~R., {et~al.} 2014, \apj,
  795, 161

\bibitem[{Douglas {et~al.}(2019)Douglas, Curtis, Agüeros, Cargile, Brewer,
  Meibom, \& Jansen}]{Douglas_2019}
Douglas, S.~T., Curtis, J.~L., Agüeros, M.~A., {et~al.} 2019, ApJ, 879, 100

\bibitem[{Durney {et~al.}(1993)Durney, De~Young, \& Roxburgh}]{Durney1993}
Durney, B.~R., De~Young, D.~S., \& Roxburgh, I.~W. 1993, Solar Physics, 145,
  207

\bibitem[{Eaton \& Hall(1979)}]{Eaton:1979aa}
Eaton, J.~A. \& Hall, D. 1979, ApJ, 227, 907

\bibitem[{{Foreman-Mackey} {et~al.}(2013){Foreman-Mackey}, {Hogg}, {Lang}, \&
  {Goodman}}]{Foreman-Mackey2013}
{Foreman-Mackey}, D., {Hogg}, D.~W., {Lang}, D., \& {Goodman}, J. 2013, \pasp,
  125, 306

\bibitem[{{Garraffo} {et~al.}(2018){Garraffo}, {Drake}, {Dotter}, {Choi},
  {Burke}, {Moschou}, {Alvarado-G{\'o}mez}, {Kashyap}, \&
  {Cohen}}]{Garraffo2018}
{Garraffo}, C., {Drake}, J.~J., {Dotter}, A., {et~al.} 2018, \apj, 862, 90

\bibitem[{{Gonz\'alez-\'Alvarez} {et~al.}(2019){Gonz\'alez-\'Alvarez}, {Micela,
  G.}, {Maldonado, J.}, {Affer, L.}, {Maggio, A.}, {Lanza, A. F.}, {Covino,
  E.}, {Benatti, S.}, {Bignamini, A.}, {Cosentino, R.}, {Damasso, M.},
  {Desidera, S.}, {Gonz\'alez Hern\'andez, J. I.}, {Mart\'{\i}nez-Fiorenzano,
  A.}, {Pagano, I.}, {Perger, M.}, {Piotto, G.}, {Pinamonti, M.}, {Rainer, M.},
  {Rebolo, R.}, {Ribas, I.}, {Scandariato, G.}, {Sozzetti, A.}, {Su\'arez
  Mascare\~no, A.}, \& {Toledo-Padr\'on, B.}}]{Gonzalez-Alvarez2019}
{Gonz\'alez-\'Alvarez}, {Micela, G.}, {Maldonado, J.}, {et~al.} 2019, A\&A,
  624, A27

\bibitem[{G{\"u}del(2004)}]{Guedel2004}
G{\"u}del, M. 2004, Astron. Astrophys. Rev., 12, 71

\bibitem[{{Irwin} {et~al.}(2011){Irwin}, {Berta}, {Burke}, {Charbonneau},
  {Nutzman}, {West}, \& {Falco}}]{Irwin2011}
{Irwin}, J., {Berta}, Z.~K., {Burke}, C.~J., {et~al.} 2011, \apj, 727, 56

\bibitem[{Jao {et~al.}(2018)Jao, Henry, Gies, \& Hambly}]{Jao_2018}
Jao, W.-C., Henry, T.~J., Gies, D.~R., \& Hambly, N.~C. 2018, ApJ, 861, L11

\bibitem[{{Kawaler}(1988)}]{Kawaler_1988}
{Kawaler}, S.~D. 1988, \apj, 333, 236

\bibitem[{L{\'{e}}pine \& Gaidos(2011)}]{L_pine_2011}
L{\'{e}}pine, S. \& Gaidos, E. 2011, AJ, 142, 138

\bibitem[{Lindegren {et~al.}(2018)Lindegren, Hern{\'{a}}ndez, Bombrun, Klioner,
  Bastian, Ramos-Lerate, de~Torres, Steidelm{\"{u}}ller, Stephenson, Hobbs,
  Lammers, Biermann, Geyer, Hilger, Michalik, Stampa, McMillan,
  Casta{\~{n}}eda, Clotet, Comoretto, Davidson, Fabricius, Gracia, Hambly,
  Hutton, Mora, Portell, van Leeuwen, Abbas, Abreu, Altmann, Andrei, Anglada,
  Balaguer-N{\'{u}}{\~{n}}ez, Barache, Becciani, Bertone, Bianchi, Bouquillon,
  Bourda, Br{\"{u}}semeister, Bucciarelli, Busonero, Buzzi, Cancelliere,
  Carlucci, Charlot, Cheek, Crosta, Crowley, de~Bruijne, de~Felice, Drimmel,
  Esquej, Fienga, Fraile, Gai, Garralda, Gonz{\'{a}}lez-Vidal, Guerra, Hauser,
  Hofmann, Holl, Jordan, Lattanzi, Lenhardt, Liao, Licata, Lister,
  L{\"{o}}ffler, Marchant, Martin-Fleitas, Messineo, Mignard, Morbidelli,
  Poggio, Riva, Rowell, Salguero, Sarasso, Sciacca, Siddiqui, Smart, Spagna,
  Steele, Taris, Torra, van Elteren, van Reeven, \& Vecchiato}]{Lindegren2018}
Lindegren, L., Hern{\'{a}}ndez, J., Bombrun, A., {et~al.} 2018, A{\&}A, 616, A2

\bibitem[{Mann {et~al.}(2015)Mann, Feiden, Gaidos, Boyajian, \&
  Braun}]{Mann2015}
Mann, A.~W., Feiden, G.~A., Gaidos, E., Boyajian, T., \& Braun, K.~V. 2015,
  ApJ, 804, 1

\bibitem[{Mann {et~al.}(2016)Mann, Feiden, Gaidos, Boyajian, \& von
  Braun}]{Mann_2016}
Mann, A.~W., Feiden, G.~A., Gaidos, E., Boyajian, T., \& von Braun, K. 2016,
  \apj, 819, 87

\bibitem[{Matt {et~al.}(2015)Matt, Brun, Baraffe, Bouvier, \&
  Chabrier}]{Matt_2015}
Matt, S.~P., Brun, A.~S., Baraffe, I., Bouvier, J., \& Chabrier, G. 2015, ApJ,
  799, L23

\bibitem[{{McQuillan} {et~al.}(2014){McQuillan}, {Mazeh}, \&
  {Aigrain}}]{Mcquillan2014}
{McQuillan}, A., {Mazeh}, T., \& {Aigrain}, S. 2014, \apjs, 211, 24

\bibitem[{{Metcalfe} \& {Egeland}(2019)}]{Metcalfe2019}
{Metcalfe}, T.~S. \& {Egeland}, R. 2019, \apj, 871, 39

\bibitem[{{Mohanty} {et~al.}(2002){Mohanty}, {Basri}, {Shu}, {Allard}, \&
  {Chabrier}}]{Mohanty2002}
{Mohanty}, S., {Basri}, G., {Shu}, F., {Allard}, F., \& {Chabrier}, G. 2002,
  \apj, 571, 469

\bibitem[{Newton {et~al.}(2017)Newton, Irwin, Charbonneau, Berlind, Calkins, \&
  Mink}]{Newton2017}
Newton, E.~R., Irwin, J., Charbonneau, D., {et~al.} 2017, \apj, 834, 85

\bibitem[{{N{\'u}{\~n}ez} {et~al.}(2015){N{\'u}{\~n}ez}, {Ag{\"u}eros},
  {Covey}, {Hartman}, {Kraus}, {Bowsher}, {Douglas}, {L{\'o}pez-Morales},
  {Pooley}, {Posselt}, {Saar}, \& {West}}]{Nunez2015}
{N{\'u}{\~n}ez}, A., {Ag{\"u}eros}, M.~A., {Covey}, K.~R., {et~al.} 2015, \apj,
  809, 161

\bibitem[{{N{\'u}{\~n}ez} {et~al.}(2017){N{\'u}{\~n}ez}, {Ag{\"u}eros},
  {Covey}, \& {L{\'o}pez-Morales}}]{Nunez2017}
{N{\'u}{\~n}ez}, A., {Ag{\"u}eros}, M.~A., {Covey}, K.~R., \&
  {L{\'o}pez-Morales}, M. 2017, \apj, 834, 176

\bibitem[{{Pallavicini} {et~al.}(1981){Pallavicini}, {Golub}, {Rosner},
  {Vaiana}, {Ayres}, \& {Linsky}}]{Pallavicini1981}
{Pallavicini}, R., {Golub}, L., {Rosner}, R., {et~al.} 1981, \apj, 248, 279

\bibitem[{Parker(1955)}]{Parker:1955aa}
Parker, E.~N. 1955, ApJ, 122, 293

\bibitem[{Pizzolato {et~al.}(2003)Pizzolato, Maggio, Micela, Sciortino, \&
  Ventura}]{Pizzolato2003}
Pizzolato, N., Maggio, A., Micela, G., Sciortino, S., \& Ventura, P. 2003,
  A{\&}A, 397, 147

\bibitem[{{Preibisch} \& {Feigelson}(2005)}]{Preibisch2005}
{Preibisch}, T. \& {Feigelson}, E.~D. 2005, \apjs, 160, 390

\bibitem[{{Raetz} {et~al.}(2020){Raetz}, {Stelzer}, {Damasso}, \&
  {Scholz}}]{Raetz2020}
{Raetz}, S., {Stelzer}, B., {Damasso}, M., \& {Scholz}, A. 2020, arXiv
  e-prints, arXiv:2003.11937

\bibitem[{Reiners {et~al.}(2014)Reiners, Sch{\"{u}}ssler, \&
  Passegger}]{Reiners2014}
Reiners, A., Sch{\"{u}}ssler, M., \& Passegger, V.~M. 2014, ApJ, 794

\bibitem[{Rosen(2016)}]{Rosen2016}
Rosen, S. 2016, VizieR Online Data Catalog, IX/47

\bibitem[{Stelzer \& Neuh\"auser(2001)}]{Stelzer2001}
Stelzer \& Neuh\"auser. 2001, A\&A, 377, 538

\bibitem[{Stelzer {et~al.}(2016)Stelzer, Damasso, Scholz, \&
  Matt}]{Stelzer2016}
Stelzer, B., Damasso, M., Scholz, A., \& Matt, S.~P. 2016, MNRAS, 463, 1844

\bibitem[{Stelzer {et~al.}(2013)Stelzer, Marino, Micela, L{\'{o}}pez-Santiago,
  \& Liefke}]{Stelzer2013}
Stelzer, B., Marino, A., Micela, G., L{\'{o}}pez-Santiago, J., \& Liefke, C.
  2013, MNRAS, 431, 2063

\bibitem[{{Vanderburg} \& {Johnson}(2014)}]{2014PASP..126..948V}
{Vanderburg}, A. \& {Johnson}, J.~A. 2014, \pasp, 126, 948

\bibitem[{Veyette \& Muirhead(2018)}]{Veyette2018}
Veyette, M.~J. \& Muirhead, P.~S. 2018, ApJ, 863, 166

\bibitem[{{Voges} {et~al.}(1999){Voges}, {Aschenbach}, {Boller},
  {Br{\"a}uninger}, {Briel}, {Burkert}, {Dennerl}, {Englhauser}, {Gruber},
  {Haberl}, {Hartner}, {Hasinger}, {K{\"u}rster}, {Pfeffermann}, {Pietsch},
  {Predehl}, {Rosso}, {Schmitt}, {Tr{\"u}mper}, \& {Zimmermann}}]{Voges1999}
{Voges}, W., {Aschenbach}, B., {Boller}, T., {et~al.} 1999, \aap, 349, 389

\bibitem[{{Voges} {et~al.}(2000){Voges}, {Aschenbach}, {Boller}, {Brauninger},
  {Briel}, {Burkert}, {Dennerl}, {Englhauser}, {Gruber}, {Haberl}, {Hartner},
  {Hasinger}, {Pfeffermann}, {Pietsch}, {Predehl}, {Schmitt}, {Trumper}, \&
  {Zimmermann}}]{Voges2000}
{Voges}, W., {Aschenbach}, B., {Boller}, T., {et~al.} 2000, \iaucirc, 7432, 3

\bibitem[{Wright \& Drake(2016)}]{Wright2016}
Wright, N.~J. \& Drake, J.~J. 2016, Nature, 535, 526

\bibitem[{Wright {et~al.}(2011)Wright, Drake, Mamajek, \& Henry}]{Wright2011}
Wright, N.~J., Drake, J.~J., Mamajek, E.~E., \& Henry, G.~W. 2011, ApJ, 743

\bibitem[{Wright {et~al.}(2018)Wright, Newton, Williams, Drake, \&
  Yadav}]{Wright2018}
Wright, N.~J., Newton, E.~R., Williams, P.~K., Drake, J.~J., \& Yadav, R.~K.
  2018, MNRAS, 479, 2351

\bibitem[{{Zechmeister} \& {K{\"u}rster}(2009)}]{2009A&A...496..577Z}
{Zechmeister}, M. \& {K{\"u}rster}, M. 2009, \aap, 496, 577

\end{thebibliography}
        
\begin{appendix}
       \renewcommand{\thetable}{A.\arabic{table}}   
       
       \onecolumn
       \setcounter{table}{0}
       \small\addtolength{\tabcolsep}{-4pt}
       
     \begin{sidewaystable}[p]\small
       		\renewcommand{\tabularxcolumn}[1]{>{\arraybackslash}m{#1}} \newcolumntype{W}{>{\centering\arraybackslash}X} 
       		\captionsetup{width=0.9\textwidth}
       		
       		\caption{Stellar parameters and updated X-ray results we computed for the 288 stars we took from the literature samples \citep{Wright2011,Wright2016,Wright2018,Stelzer2016,Gonzalez-Alvarez2019}. The last column shows the $\rm FLAG$ we used to verify if {\em Gaia} distances are reliable as  explained in Sect.~\ref{stellar_par}}.
       		 \label{ALLdata_par}
       	\begin{center}
       		\begin{tabularx}{0.9\textheight}{lccccccccccccc}
       		
       		\toprule[0.5mm]
       		\multicolumn{1}{c}{\textbf{Name}}  
       		&\multicolumn{1}{c}{\textbf{$M_{\rm K_{s}}$}}  
       		&\multicolumn{1}{c}{\textbf{$M_{\star}$}}  
       		&\multicolumn{1}{c}{\textbf{$R_{\star}$}} 
       		&\multicolumn{1}{c}{\textbf{$\logten\left(\frac{L_{\rm bol}}{L_{\odot}}\right)$}}
       		&\multicolumn{1}{c}{\textbf{$V-J$}}
       		&\multicolumn{1}{c}{\textbf{$P_{\rm rot}$}}
       		&\multicolumn{1}{c}{\textbf{$\log\left(L_{\rm x}\right)$}}
       		&\multicolumn{1}{c}{\textbf{$\log\left(\frac{L_{\rm x}}{L_{\rm bol}}\right)$}}
       		&\multicolumn{1}{c}{\textbf{$R_{\rm O,C\&S}$}}
       		&\multicolumn{1}{c}{\textbf{$R_{\rm O,B}$}}
       		&\multicolumn{1}{c}{\textbf{$R_{\rm O,W}$}}
       		&\multicolumn{1}{c}{\textbf{$D$}}
       		&\multicolumn{1}{c}{\textbf{$FLAG_{\rm GAIA}$}}\\
       		
       		\multicolumn{1}{c}{} 
       		&\multicolumn{1}{c}{\textbf{[mag]}}  
       		&\multicolumn{1}{c}{\textbf{[$\rm M_{\odot}$]}}  
       		&\multicolumn{1}{c}{\textbf{[$\rm R_{\odot}$]}} 
       		&\multicolumn{1}{c}{} 
       		&\multicolumn{1}{c}{\textbf{[mag]}}
       		&\multicolumn{1}{c}{\textbf{[d]}}
       		&\multicolumn{1}{c}{\textbf{[erg/s]}}
       		&\multicolumn{1}{c}{}
       		&\multicolumn{1}{c}{}
       		&\multicolumn{1}{c}{}
       		&\multicolumn{1}{c}{}
       		&\multicolumn{1}{c}{\textbf{[pc]}}
       		&\multicolumn{1}{c}{}\\
       		\midrule[0.5mm] 
       		HAT 122$-$01032&6.33$\pm$0.03&0.41$\pm$0.01&0.40$\pm$0.01&-1.73$\pm$0.04&4.26&\phantom{00}4.38&28.91$\pm$0.01&-2.94$\pm$0.09&---&---&0.05&\phantom{0}18.63$\pm$0.03&1 1\\
       		LP 149$-$56&5.85$\pm$0.03&0.49$\pm$0.01&0.47$\pm$0.01&-1.51$\pm$0.04&3.71&\phantom{00}6.17&28.84$\pm$0.03&-3.24$\pm$0.10&0.14&0.25&0.10&\phantom{0}29.56$\pm$0.05&1 1\\
       		G 172$-$1&7.14$\pm$0.10&0.30$\pm$0.01&0.29$\pm$0.01&-2.06$\pm$0.19&4.32&\phantom{00}1.09&28.72$\pm$0.04&-2.81$\pm$0.23&---&---&0.01&\phantom{0}14.77$\pm$0.71&0 1\\
       		UCAC4 729$-$006249&5.14$\pm$0.03&0.61$\pm$0.01&0.59$\pm$0.02&-1.16$\pm$0.03&2.80&\phantom{00}8.35&29.09$\pm$0.03&-3.33$\pm$0.10&0.24&0.45&0.23&\phantom{0}43.63$\pm$0.07&1 1\\
       		2MASS J00380001$+$4353454&4.24$\pm$0.16&0.76$\pm$0.03&0.76$\pm$0.04&-0.81$\pm$0.12&2.95&\phantom{00}0.55&29.91$\pm$0.03&-2.87$\pm$0.34&0.01&0.07&0.01&\phantom{0}97.19$\pm$1.32&1 1\\
       		1RXS J003926.5$+$381607&4.48$\pm$0.16&0.72$\pm$0.03&0.71$\pm$0.04&-0.89$\pm$0.13&2.77&\phantom{00}3.08&29.36$\pm$0.12&-3.33$\pm$0.35&0.09&0.26&0.09&115.40$\pm$1.59&1 1\\
       		\dots&\dots&\dots&\dots&\dots&\dots&\dots&\dots&\dots&\dots&\dots&\dots&\dots&\dots\\
       		\dots&\dots&\dots&\dots&\dots&\dots&\dots&\dots&\dots&\dots&\dots&\dots&\dots&\dots\\
       		   
        \bottomrule     
        
  		\end{tabularx}  
  \begin{tablenotes}
  	\item[1] \hspace{0.8cm}\textbf{Note}: The full table is available in electronic form at the CDS.
  \end{tablenotes}
	\end{center}
 \end{sidewaystable}	   

        \renewcommand\thefigure{A.\arabic{figure}}    

                \begin{figure*}
                        \centering
                        \begin{minipage}{0.5\textwidth}
                        \centering
                        \includegraphics[width=\textwidth]{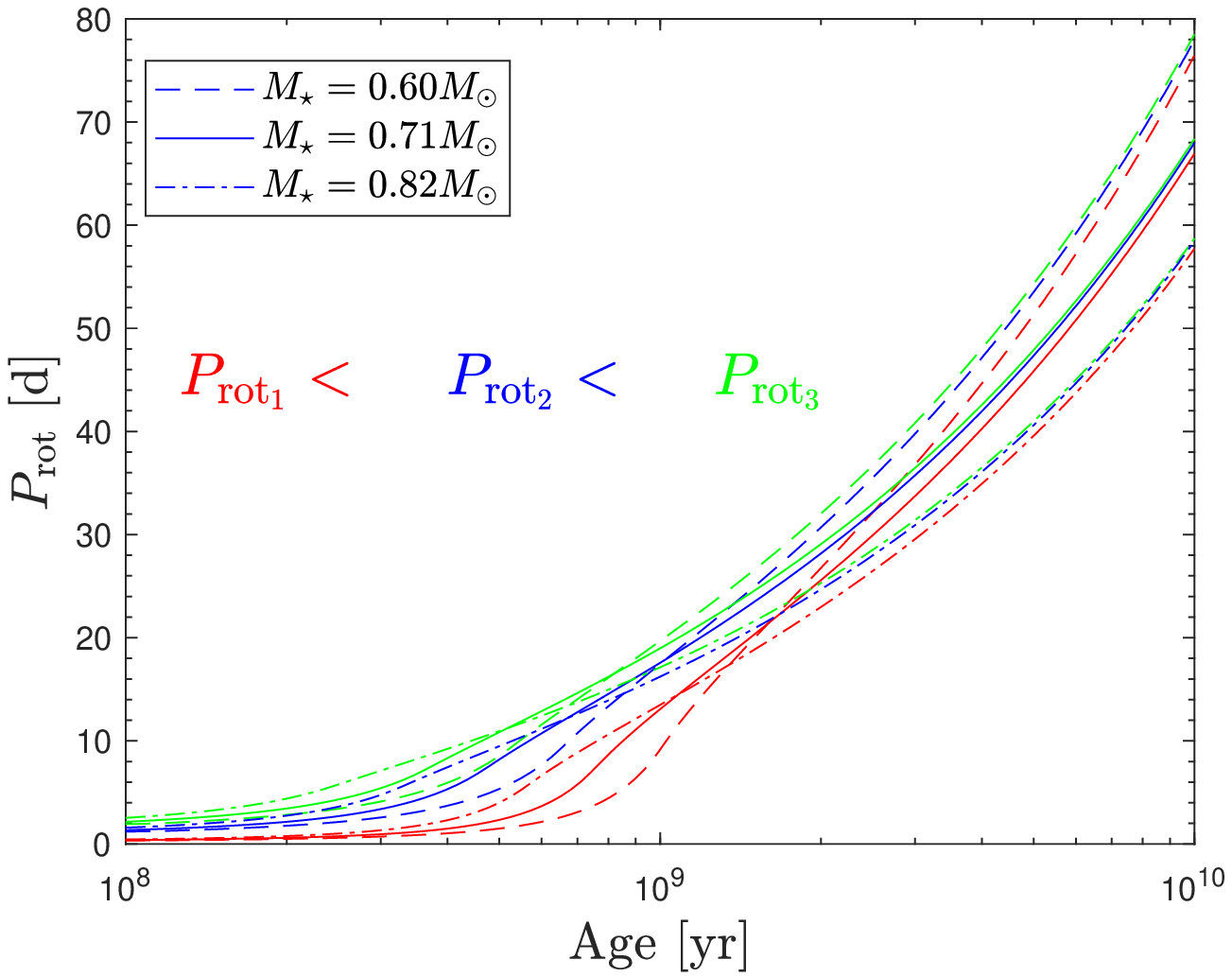}
                        \end{minipage}%
                        \begin{minipage}{0.5\textwidth}
                        \centering
                        \includegraphics[width=\textwidth]{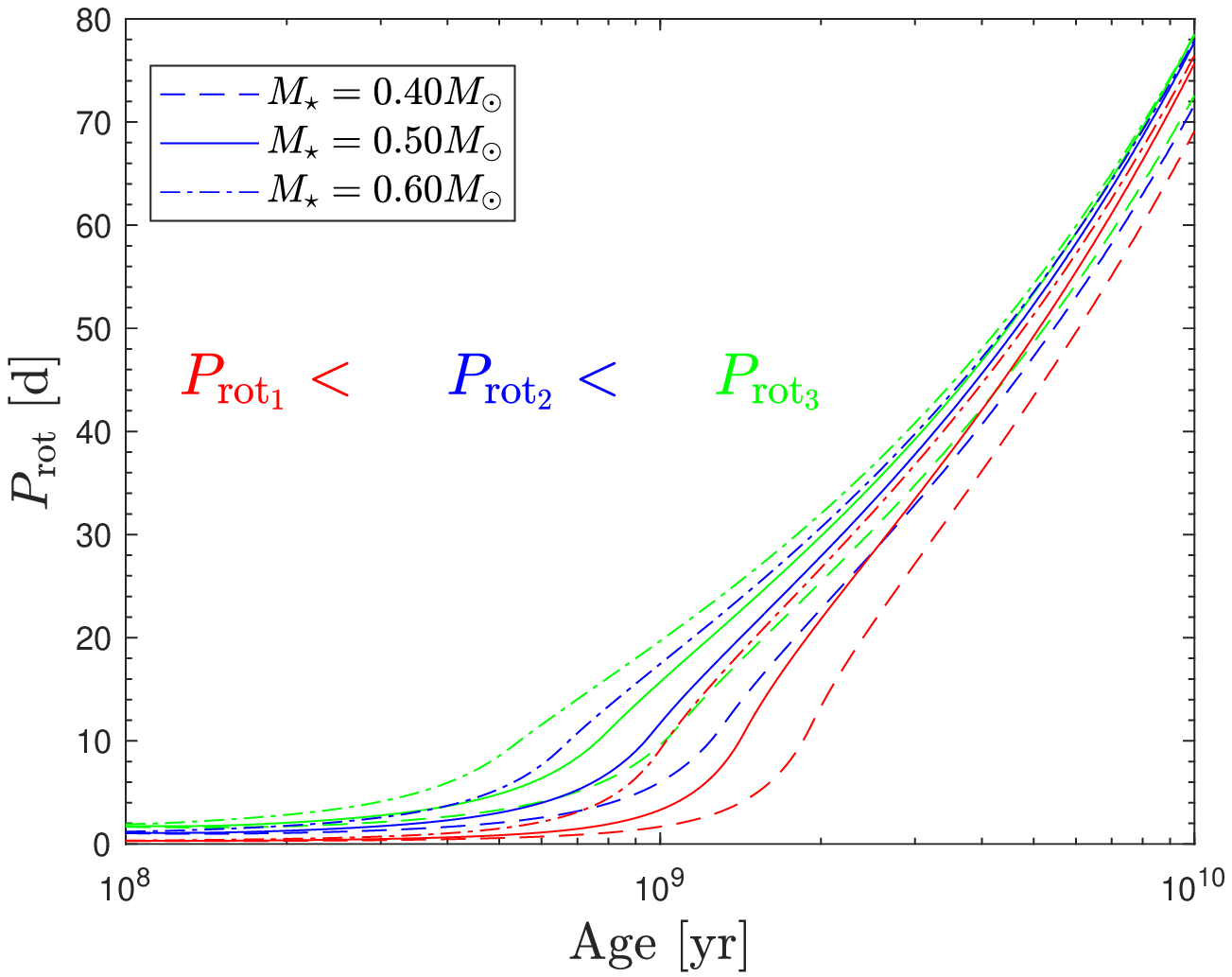}
                        \end{minipage}
                        \begin{minipage}{0.5\textwidth}
                        \centering
                        \includegraphics[width=\textwidth]{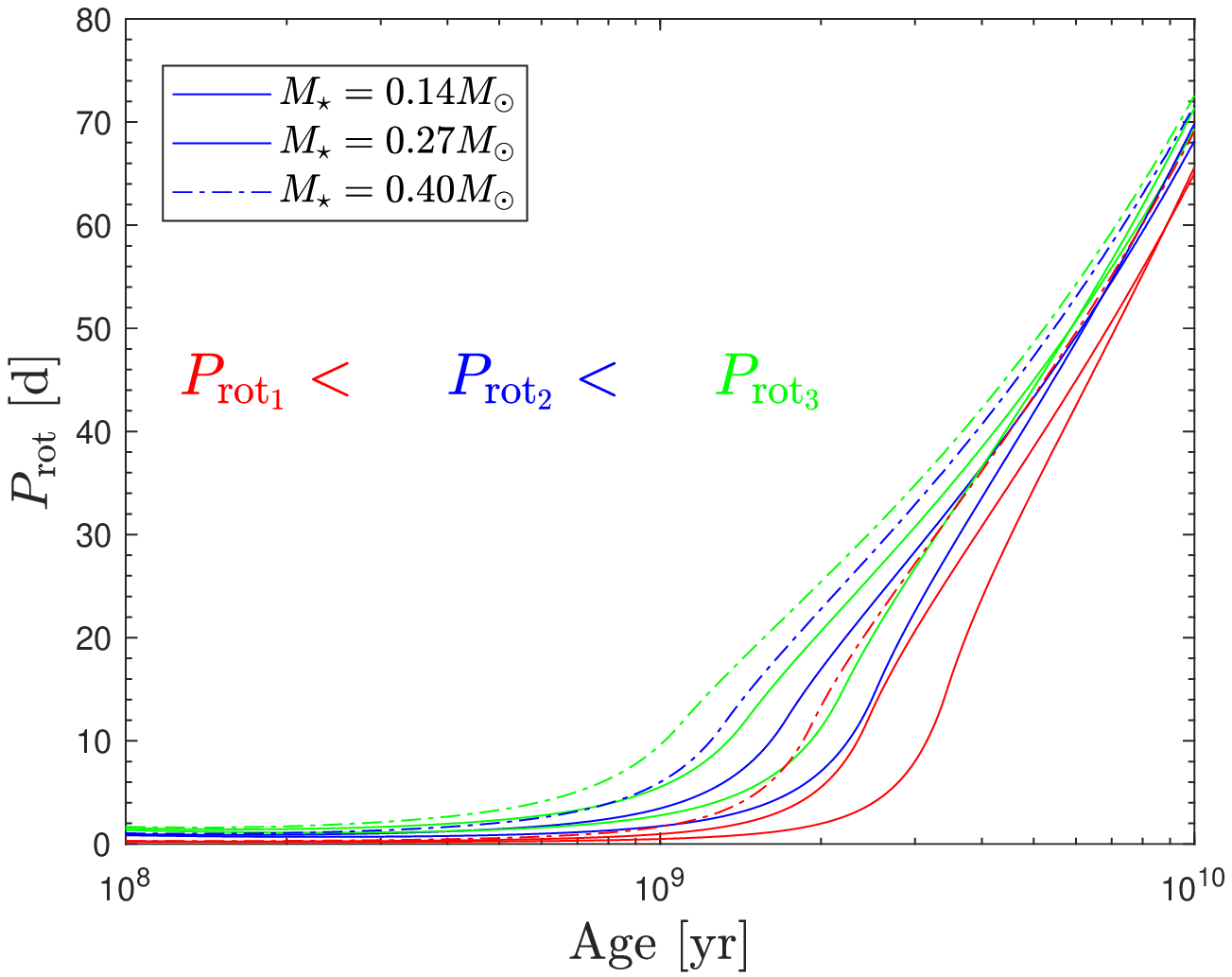}
                        \end{minipage}
                        \caption{The time-evolution models by \citet{Matt_2015} for three different initial rotation periods. In particolar, $P_{\rm rot_{1}} = 1.5454$~d, $P_{\rm rot_{2}}=5.51$~d and $P_{\rm rot_{3}}=8.83$~d.The retrieved $L_{\rm x}$-Age relation from time evolution models together with literature data by \citet{Stelzer2001}, \citet{Wright2011}, and \citet{Veyette2018}.
                        The three mass ranges are showed from the \textbf{From the left to the top}: $M_{\star} > 0.6\hspace{0.5mm}M_{\odot}$, $0.4\hspace{0.5mm}M_{\odot} \leq M_{\star} \leq 0.6\hspace{0.5mm}M_{\odot}$, and $M_{\star} < 0.4 M_{\odot}$.}
                        \label{ProtAge}
                \end{figure*}
        \end{appendix}
\end{document}